\documentclass[a4paper,11pt]{article}%
\usepackage{amssymb}
\usepackage{graphicx}
\usepackage{amsbsy}
\usepackage{amsmath}
\usepackage{cite}
\usepackage{slashed}
\usepackage{amsfonts}
\usepackage{hyperref}
\usepackage{comment}
\usepackage{braket}
\usepackage{tensor}

\usepackage[utf8]{inputenc}

\usepackage{tikz}
\usetikzlibrary{shapes,arrows,positioning,automata,backgrounds,calc,er,patterns}
\usepackage[compat=1.1.0]{tikz-feynman} 

\usepackage{listings}
\usepackage{color}

\definecolor{mygreen}{rgb}{0,0.6,0}
\definecolor{mygray}{rgb}{0.5,0.5,0.5}
\definecolor{mymauve}{rgb}{0.58,0,0.82}

\definecolor{darkWhite}{rgb}{0.94,0.94,0.94}

\newcommand{\dd} {\mathrm{d}}

\newcommand{\tr} {\mathrm{tr}}
\newcommand{\Tr} {\mathrm{Tr}}
\newcommand{\boundellipse}[3]
{(#1) ellipse (#2 and #3)
}
\newcommand{\dc} {\mathcal{D}}

\newcommand{\g}{\sqrt{-g}}

\newcommand{\sP}{\slashed{P}}
\newcommand{\sD}{\slashed{D}}

\DeclareMathOperator{\pa}{\partial}
\DeclareMathOperator{\half}{\frac{1}{2}}

\newcommand\T{\rule{0pt}{3.2ex}}
\newcommand\B{\rule[-2.1ex]{0pt}{0pt}}

\usepackage{dsfont}

\def\logm#1{\log\frac{#1}{\mu^2}}

\newcommand{\Integral}{\mathcal{I}}

\usepackage{multicol}
\usepackage{color}
\definecolor{verdes}{cmyk}{0.92,0,0.59,0.4}

\definecolor{Grn}{rgb}{0.1,0.5,0.2}
\definecolor{Blu}{rgb}{0.,0.,1.}
\definecolor{Red}{rgb}{0.7,0.1,0.1}
\definecolor{SE}{rgb}{0.5,0,0.4}
\definecolor{Tur}{rgb}{0,0.75,0.65}

\newcommand{\Blu}[1]{{\color{Blu}{#1}}} 
 
\newcommand{\SE}[1]{{\color{SE}{#1}}}

\newcommand{\rl}[1]{\SE{[Remy:\,#1]}} 
         
\newcommand{\jq}[1]{\Blu{[Jeremie:\,#1]}}

\usepackage{hyperref} 
\hypersetup{
     colorlinks = true,
     linkcolor = blue,
     citecolor = magenta,
     anchorcolor = blue,
     filecolor = blue,
     urlcolor = magenta,
     bookmarks=true,
     linktocpage =true,
     }

\makeatletter
\renewcommand\@dotsep{200}
\makeatother

\begin{document}
\thispagestyle{empty} \setcounter{page}{0} \begin{flushright}
May 2022\\
\end{flushright}

\vskip          4.1 true cm

\begin{center}
{\huge Anomalies from an effective field theory perspective}
\\[1.9cm]

\textsc{Baptiste Filoche}$^{a,b,c}$, \textsc{R\'emy Larue}$^{a,d}$, \textsc{J\'er\'emie Quevillon}$^{a}$ \\ \textsc{\ and Pham Ngoc Hoa Vuong}$^{a}$%
\vspace{0.5cm}\\[9pt]

	{$^a$\it Laboratoire de Physique Subatomique et de Cosmologie,}\\
	{\it Universit\'{e} Grenoble-Alpes, CNRS/IN2P3, Grenoble INP, 38000 Grenoble, France}\\[2mm]
	{$^b$ \it ENS de Lyon, Université Claude Bernard, 69342 Lyon, France}\\[2mm]
	{$^c$ \it Deutsches Elektronen-Synchrotron DESY,}\\
	{\it Notkestr. 85, 22607 Hamburg, Germany}\\[2mm]
	{$^d$ \it ENS Paris-Saclay, }\\
	{\it 4, avenue des Sciences
91190 Gif-sur-Yvette, France}\\[2mm]
	
	\vspace{2cm}
		
\textbf{Abstract}\smallskip
\end{center}

The path-integral measure of a gauge-invariant fermion theory is transformed under the chiral transformation and leads to an elegant derivation of the anomalous chiral Ward-Takahashi identities, as we know from the seminal work of Fujikawa. We present in this work an alternative and illuminating way to calculate the Jacobian in the path-integral measure from the Covariant Derivative Expansion technique used in Effective Field Theory. We present several ways to customise the crucial regularisation such that the anomaly is located in the desired current, which is unprecedented within the path integral approach. We are then able to derive, in a transparent and unified way the covariant, consistent, gravitational and scale anomalies.

\newpage

\setcounter{tocdepth}{3}
\tableofcontents

\newpage

\section{Introduction}

Symmetries play an important role in explaining the fundamental forces of nature. A symmetry valid in the classical theory might be violated in its quantised version. This defines what an anomaly is in Quantum Field Theory (QFT). The axial or chiral anomaly which has a long history is certainly the most well-known and had a huge impact in the building and understanding of QFT.

In 1967, Sutherland and Veltman \cite{Sutherland:1967vf,Veltman:1967vf} proved that the neutral pion, $\pi_0$, cannot decay into two photons in obvious disagreement with the experimental results. The $\pi_0 \rightarrow \gamma\gamma$ puzzle has been solved in 1969 by Bell and Jackiw \cite{Bell:1969ts} who showed that the, unexpected, axial symmetry breaking perfectly explains this decay, later confirmed by Alder \cite{Adler:1969gk}. This is the so called ABJ anomaly now commonly computed through triangle Feynman diagrams involving one axial and two vector currents and involving a UV divergence which leads to $\partial^{\mu}j^{5}_{\mu}=1/(8\pi^2) F\tilde{F}$, meaning that while the vector conservation law can be maintained, the axial current has to be broken.

As stated by the Adler-Bardeen theorem \cite{Adler:1969er}, this is actually quite astonishing that the anomaly does not receive radiative corrections and is totally given at the one-loop level.
It has been realised later~\cite{Dolgov:1971iz} that the anomaly was not just a perturbation effect arising from divergent diagrams requiring to be regularised. Indeed, anomalies, as opposed to divergences, essentially do not diverge even if they both emerge from the presence of an infinite number of degrees of freedom in the theory~\footnote{In that regard, the scale anomaly is singular.}. It seems more accurate to appreciate anomalies as a side effect of the quantisation which might break some symmetries.

This is really in the seventies~\cite{Jackiw:1976pf,Nielsen:1976hs,Nielsen:1977aw,Nielsen:1977qm} that the anomaly was interpreted in term of a topological invariant. Anomaly has been indeed determined by an index theorem by counting the zero-modes of a chiral Dirac operator. This counting was made transparent by Fujikawa~\cite{Fujikawa:2004cx} as the anomaly arises in the path integral as the functional trace of $\gamma_5$.

In QFT the fundamental quantity is the generating functional  which is a path integral for the classical action. How can an anomaly emerge when the classical action is invariant under a symmetry? This questions has been solved by Fujikawa in Ref.~\cite{Fujikawa:1979ay,Fujikawa:1980eg,Fujikawa:1983bg} by realising that the only quantity which contains the quantum aspects, the path integral measure, does not remain invariant under chiral transformations. The anomaly is precisely arising from the associated non-trivial Jacobian, which is ill-defined. To regularise it in a gauge-invariant manner, one can use, as Fujikawa did, an eigenbasis expansion associated to a Gaussian cutoff or alternatively a heat-kernel regularisation or a $\zeta$–function regularisation\footnote{See Ref.~\cite{Bertlmann:1996xk, Fujikawa:2004cx} and the references therein.}. In any case, the anomaly technically arises as a finite term from the regularisation. Within this formalism the anomaly is truly independent of perturbation theory and indeed provides a conceptually and satisfactory derivation of the anomaly terms present from the beginning instead of discovering it after the evaluation of the divergence of a current.\\

In particle physics, the methods of Effective Field Theory (EFT) have recently seen a resurgence, mostly due to the lack of new physics discovery at the weak scale. Observations seem to indicate that new physics should indeed be decoupled to heavier scales, urging us to reconsider the Standard Model (SM) as a more humble EFT supplemented by higher-dimensional operators. 

The new physics integrated out at some higher energy scale is technically encapsulated in the coefficients of these higher-dimensional operators. The task to evaluate these Wilson coefficients from ultraviolet (UV) theories has traditionally been done using Feynman diagrams, where amplitudes involving the heavy degrees of freedom are explicitly “matched” to the EFT amplitudes. However, a more elegant approach is to “integrate out” the heavy particles by evaluating the path integral directly~\cite{Gaillard:1985uh,Cheyette:1987qz,Henning:2014wua,Drozd:2015rsp} even if in the past, this approach has been limited because, in practice, the expansion techniques could be cumbersome. However, recently a significant effort has been done for developing new methods to evaluate the path integral at one loop more efficiently using improved expansion techniques~\cite{Fuentes-Martin:2016uol,Zhang:2016pja,Ellis:2017jns,Ellis:2020ivx,Cohen:2020fcu}.

In this work, we propose to compute anomalies in QFT, identified as a Jacobian in the path integral formalism as Fujikawa did, but in view of recent developments made in EFTs and more especially the usefulness of a mass expansion technique such as the Covariant Derivative Expansion (CDE)~\cite{Gaillard:1985uh,Cheyette:1987qz,Henning:2014wua,Drozd:2015rsp}.
This offers a novel technical approach to evaluate anomalies in QFT within the path integral formalism. The novelty of our formalism is the following. First, it does not truly rely on the computation of the transformation of the measure through the existence and definition of the Dirac operator spectrum and more especially trying to properly deal with the zero modes of the chiral Dirac operators as Fujikawa did. Second, the anomalies emerge from a ratio of two ill-defined determinants which can be evaluated systematically and efficiently by the CDE technique.

In practice, in Fujikawa's method, the various symmetries have been in-forced to the model beforehand in order to define the eigenbasis of the Dirac operator and cure the illness of the Jacobian of the considered transformation. The choice of regulator (to count the zero modes) to evaluate the anomaly is crucial and depends on the active symmetries. It will lead to several type of anomalies (consistent, covariant, etc.). 
We will see that in many situations, it is possible to end-up to this situation when ``bosonising'' the fermionic functional determinants, then straightforwardly extracting the anomalous interactions with the CDE. Within our proposed alternative method, the regularisation procedure is fixed and always carried with the usual dimensional-regularisation scheme~\cite{tHooft:1972tcz}. The illness of the Jacobian is then embodied in the ambiguity of Dirac traces involving $\gamma_5$ (see Ref.~\cite{Elias:1982ea,Quevillon:2021sfz}). These ambiguities are cured by imposing manually the invariance of the EFT under specific symmetries. Thus, our method is available to evaluate in a general way the covariant and consistent anomaly from the path integral having then the possibility to tune which current bears the anomaly.

These two approaches allow to treat gauge and mixed global-gauge anomalies i.e consistent and covariant anomalies, gravitational anomaly, as well as the scale anomaly, in a transparent, simple and unified way which certainly deserves to be presented due to the importance and phenomenological implications of anomalies in physics.\\

The plan of the paper is the following, in the next section we detail the outline of the proposed new method to compute QFT anomalies within the path integral formalism and use the example of the axial anomaly to concretely show how to connect the Jacobian of the transformation, to the functional determinants and how to conveniently expand them. In a third section, we apply our formalism to other anomalous transformations in chiral gauge field theory, namely fermionic vector and axial transformations, leading to so-called covariant and consistent anomalies. In the fourth section, we evaluate the axial-gravitational anomaly and technically show how to deal with this approach in curved space-time. In the fifth section, we evaluate the so-called scale anomaly, without having to introduce the curvature of space-time~\cite{Coleman:1970je, Crewther:1972kn, Fujikawa:1980vr, PhysRevD.23.2262, Adler:1976zt}. In a subsequent section, we discuss the approach of the new method presented in this work compared to the original approach of Fujikawa, before bringing our conclusion in the last section.

\section{Outline of the new method}\label{sec:Main}

In this section we will introduce and present a method to compute QFT anomalies within the path integral formalism while dealing with EFTs. In order to make our points as clear as possible we will deal with the concrete case of the axial anomaly in a vector gauge theory. More general situations will be discussed in the following section.

\subsection{Functional determinant and Jacobian}

Let us start with a Dirac fermion field involved in a vector gauge theory with the following path integral,
\begin{equation}
	Z \equiv \int \dc \psi  \dc \bar \psi \exp \left(\ i\int \dd^4 x \bar \psi (i \slashed \partial - \slashed V - m) \psi \right) = \int \dc \psi \dc \bar \psi e^{iS}, \label{Zini}
\end{equation}
with $V$ a gauge field, element of $SU(N)\equiv G$ and ``slashed'' quantities are Lorentz-contracted with $\gamma$ matrices. Performing the integration on Grassmann variables $Z$ can be written as (in the eigenbasis with eigenvalues $\lambda_n$ of the Dirac operator),
\begin{equation}
	Z = \prod_n (\lambda_n - m) = \det\left(i\slashed \partial - \slashed V -m\right),
\end{equation}
where {\it det} is a functional determinant. Let us consider an infinitesimal chiral reparametrisation of the fermionic field, of parameter $\theta(x)=\theta^a T^a\in SU(N)$,
\begin{equation}
	\psi \to e^{i\theta(x)\gamma_5}\psi,\quad \bar \psi \to \bar \psi e^{i\theta(x)\gamma_5}\, .
	\label{FermionRep: axial}
\end{equation}
Under such a transformation, the path integral measure transforms with a Jacobian $J[\theta]$,
\begin{equation}
	\dc \psi \dc \bar \psi \to J[\theta] \dc \psi \dc \bar \psi,
\end{equation}
on the other hand the action transforms like,
\begin{equation}
	S \to S - \int \dd^4 x \, \bar \psi \bigg[ 2im \theta \gamma_5 + (\slashed D \theta) \gamma_5 \bigg] \psi,
\end{equation}
with $(\slashed D\theta) = (\slashed \partial \theta) + i[\slashed V,\theta]$ and the parenthesis indicates the local derivative. The path integral after the chiral reparametrisation reads,
\begin{equation}\label{eq:PIinv}
	Z^\prime = \int J[\theta] \dc \psi \dc \bar \psi \exp\left(iS - i\int \dd^4 x \, \bar\psi \bigg[ 2im\theta \gamma_5 + (\slashed D \theta) \gamma_5 \bigg]  \psi \right).
\end{equation}
Since the anomaly is fully determined by the structure of the gauge groups of the theory, the Jacobian $J[\theta]$ does not depend on the fermionic field, then one can perform the integration on the Grassmann variables and write,
\begin{equation}
	Z^\prime = J[\theta] \det(i\slashed \partial - \slashed V - m - 2im\theta \gamma_5 - (\slashed D \theta) \gamma_5).
\end{equation}
As a result of the invariance under the labeling of the path integral variables ($Z=Z^{\prime}$), the Jacobian reads, 
\begin{equation}
	J[\theta] = \frac{\det (i\slashed D -m)}{\det(i\slashed D - m -2im\theta \gamma_5 - (\slashed D \theta) \gamma_5)} = \frac{\det(i\slashed D -m)}{\det(i\slashed D -m + i\{\theta \gamma_5,i\slashed D -m\})} 
	\label{Jacdet} \ .
\end{equation}
The Jacobian can therefore be expressed as the exponential of the difference of two functional determinants,
\begin{equation}
	J[\theta] = \exp\left(\log\det(i\slashed D -m) - \log\det(i\slashed D - m + i\{\theta \gamma_5,i\slashed D -m\}) \right) \equiv \exp\left[\mathcal \int \dd^4 x \, \mathcal{A}(x) \right].
\end{equation}
In the peculiar case of the chiral reparametrisation of Eq.~\eqref{FermionRep: axial} being disjoint from gauge transformations, injecting this solution for $J[\theta]$ in Eq.~\eqref{eq:PIinv} leads to the relation,
\begin{equation}
    D_\mu\braket{\bar{\psi}\gamma^{\mu}\gamma_5\psi} =2im \braket{\bar \psi \gamma_5 \psi} +\frac{\delta \mathcal{A}(x)}{\delta \theta(x)}
    \,,
    \label{WI}
\end{equation}
which is the anomalous Ward identity of the axial current reflecting the anomalous behaviour of that chiral reparametrisation.\\

The main goal of this paper is to compute the anomaly operator of a theory, $\cal{A}$, directly from its path integral formulation. Yet, we will not revert to the procedure of Fujikawa to compute the determinants, which corresponds to a precise procedure to regularise the computation (the core of the problem). Instead we will call in the mass expansion method known as Covariant Derivative Expansion (CDE)~\cite{Gaillard:1985uh,Cheyette:1987qz} that we will combine with different regularisation procedures. All in all, being very efficient to obtain anomalies in QFT.

One should also note that the CDE method has recently proved its usefulness while dealing with precisely this kind of EFTs and more especially the {\it matching} step which consists in expressing the Wilson coefficient of the low energy EFT as a function of the parameters of the high energy theory (see for example Refs.~\cite{Henning:2014wua,Drozd:2015rsp,Ellis:2016enq,Henning:2016lyp,Ellis:2017jns,Ellis:2020ivx,Cohen:2020fcu}).

\subsection{The ABJ anomaly from the Covariant Derivative Expansion}

The principle of the CDE approach will be detailed below. Let $\mathcal A$ be,
\begin{equation}
	\int \dd^4 x \, \mathcal A(x) = -\Tr \log \left(i\slashed D -m -2im\theta \gamma_5 - (\slashed \partial \theta) \gamma_5\right) +\Tr \log \left(i\slashed D -m\right) \ . \label{intA}
\end{equation}
In this section, we restrain ourselves to a vector gauge theory, $D_{\mu}=\partial_{\mu}+iV_{\mu}$, with $V \in G = SU(N)$, and the chiral reparametrisation of the fermionic field is a simple axial $U(1)$ transformation. Thus we expect to obtain the so-called chiral or Adler–Bell–Jackiw (ABJ) anomaly~\cite{Adler:1969er,Bell:1969ts}.

For clarity, we will first present the evaluation of the first functional trace in Eq.~\eqref{intA} that we label $\mathcal{A}_\theta$, before combining both needed to compute the axial current anomaly ${\mathcal A}$. We evaluate the trace over space-time using a plane wave basis, leaving the trace \textit{tr} over the internal space,
\begin{equation}
	{\mathcal A}_\theta = -\int \frac{\dd^d q}{(2\pi)^d} e^{iq\cdot x} \tr \log \left(i\slashed D -m -2im\theta \gamma_5 - (\slashed \partial \theta) \gamma_5\right) e^{-iq\cdot x} \ ,
\end{equation}
use the Baker-Campbell-Hausdorff formula to perform the spatial translation,
\begin{equation}
	{\mathcal A}_\theta = -\int \frac{\dd^d q}{(2\pi)^d} \tr \log \left(i\slashed D + \slashed q -m -2im\theta \gamma_5 - (\slashed \partial \theta) \gamma_5\right) \ ,
\end{equation}
and perform the change of variable $q \to -q$ to factorise an inverse propagator-like term,
\begin{equation}
	{\mathcal A}_\theta = -\int \frac{\dd^d q}{(2\pi)^d} \tr \log \left[-(\slashed q +m)\left(1 +\frac{-1}{\slashed q + m} (i\slashed D - 2im\theta \gamma_5 - (\slashed \partial \theta) \gamma_5) \right)\right] \ .
\end{equation}
The factorised term exhibits UV divergences, and involve the
usual scale of renormalisation, which would be introduced
through dimensional regularisation. It can be absorbed in redefinitions of the parameters
of the model~\footnote{It corresponds to a renormalisation of the vacuum energy and it can be absorbed as a constant term in the Standard Model Higgs potential.}. One could also notice that it would be anyway canceled by the other trace to evaluate in Eq.~\eqref{intA}.
Using Taylor expansion on the remaining logarithm,
\begin{equation}
	{\mathcal A}_\theta = \left. \int \frac{\dd^d q}{(2\pi)^d} \sum_{n=1}^\infty \frac{1}{n} \tr \left[ \frac{-1}{\slashed q +m} \left(- i\slashed D + 2im\theta \gamma_5 + (\slashed \partial \theta) \gamma_5 \right) \right]^n \right.
	\, .
\end{equation}
If we now apply the very same treatment to the other contribution, $\Tr\log(i\sD-m)$ of Eq.~\eqref{intA}, in order to evaluate the anomaly, we find that the terms which do not involve the $\theta$ parameter do cancel with each other,
\begin{equation}
	\mathcal A = \left.\int \frac{\dd^d q}{(2\pi)^d} \sum_{n=1}^\infty \frac{1}{n} \tr \left[ \frac{-1}{\slashed q +m} \left(- i\slashed D + 2im\theta \gamma_5 + (\slashed \partial \theta) \gamma_5 \right) \right]^n \right|_{\text{carrying $\theta$ dependence}} \ .
	\label{ano4d}
\end{equation}
As shown in appendix \ref{RatioDetExp}, it can alternatively be written as,
\begin{equation}
	\mathcal A = \int \frac{\dd^d q}{(2\pi)^d}\tr\,\left(2im\theta \gamma_5 + (\slashed \partial \theta) \gamma_5\right) \sum_{n=0}^\infty \left[ \frac{-1}{\slashed q +m} \left(- i\slashed D  \right) \right]^n \frac{-1}{\slashed q +m}
	\, .
	\label{ano4d gamm5-first}
\end{equation}
So the reader should not be surprised if we switch between the two expressions. 
This expression might still look quite cumbersome to deal with, however, as we will see in the following section, it only calls for a basic power counting and use of master integrals\,\footnote{
Notice that our formalism does not support the $m=0$ case. We suggest the Heat-kernel method if the reader would like to recover the ABJ anomaly in this case.}.

\subsection{Complete evaluation of the ABJ anomaly from the CDE}

Since we are only interested in the terms linear~\footnote{This is the only possibility to obtain a $\theta$-dependent term times a gauge boundary term which is mass independent.} in $\theta$ in Eq.~\eqref{ano4d}, the anomaly can be expressed as $\mathcal A = \mathcal A^{m \gamma_5} + \mathcal A^{\slashed \partial \gamma_5}$~\footnote{This is nothing but the transcription of the Ward identity Eq.~\eqref{WI} in the present CDE context.} with,

\begin{align}
	\mathcal A^{m \gamma_5} =&\left.\int \frac{\dd^d q}{(2\pi)^d} \sum_{n=1}^\infty \frac{1}{n} \tr \left[ \frac{-1}{\slashed q +m} \left(- i\slashed D + 2im\theta \gamma_5 \right) \right]^n \right|_{\mathcal{O}(\theta)}\;, \nonumber \\
    \mathcal A^{\slashed \partial \gamma_5} =& \left.\int \frac{\dd^d q}{(2\pi)^d} \sum_{n=1}^\infty \frac{1}{n} \tr \left[ \frac{-1}{\slashed q +m} \left( - i\slashed D + (\slashed \partial \theta) \gamma_5 \right) \right]^n \right|_{\mathcal{O}(\theta)}\;.
    \label{Adecompose}
\end{align}

The terms which contribute to $\mathcal A$ involve, here, exactly one $\gamma_5$ matrix and there can be no contribution from orders greater than $n=5$, within the CDE approach, since they would carry a mass dependence.

Some of the integrals in Eq.~\eqref{Adecompose} are divergent and we use dimensional regularisation~\cite{tHooft:1972tcz} to evaluate them along with the $\overline{MS}$ scheme for renormalisation. The traces over Dirac matrices have to be performed in $d=4-\epsilon$ dimensions, and the $\epsilon$-terms resulting from the contractions with the metric tensor (satisfying then $g^{\mu\nu}g_{\mu\nu}=d$) must be kept in the calculations. These $\epsilon$-terms will then multiply with the $(1/\epsilon)$ pole of the divergent integrals and yield finite contributions. We also emphasise that depending on the regularisation scheme for $\gamma_5$ in $d$-dimensions, different results for $\epsilon$-terms in Dirac traces will emerge (see for examples Refs.~\cite{Chanowitz:1979zu,Novotny:1994yx}). In the following sections, we will discuss in details several prescriptions that one can use to evaluate ill-defined Dirac traces involving $\gamma_5$ matrices, in dimensional regularisation. However, in this section, since we discuss the case of a vector gauge theory related to Eq.~\eqref{Zini}, the divergent contribution are regularised using Breitenlohner-Maison-'t Hooft-Veltman (BMHV) scheme of dimensional regularisation~\cite{tHooft:1972tcz,Breitenlohner:1977hr} which is compatible with the conservation of the gauge vector current at the quantum level, as it is well-known, placing then the anomaly entirely in the classically conserved (only) axial current associated to Eq.~\eqref{FermionRep: axial}. In the evaluation of Eq.~\eqref{Adecompose}, we will then maintain the trace cyclicity property which might not hold for another $\gamma_5$ regularisation scheme.

We are here by-passing, on purpose, an important difficulty regarding the crucial regularising step in order to focus on the standard but important CDE algebra. The reader willing to concentrate on a careful regularisation procedure should directly reach the next section.

Regarding the actual task of collecting operators from Eq.~\eqref{Adecompose}, we do not especially rely on it but the different contributions produced by these expansions could also be enumerated using the convenient formalism of covariant diagrams (see Ref.~\cite{Zhang:2016pja} for example).

To perform the computations straightforwardly from Eq.~\eqref{Adecompose}, we decompose the propagator $-1/(\slashed{q}+m)$ as follows,
\begin{align}
    \dfrac{-1}{\slashed{q}+m} = \dfrac{m}{q^2-m^2} + \dfrac{-\slashed{q}}{q^2-m^2}
    \, .
\end{align}

Let us consider first the expansion of $\mathcal A^{\slashed \partial \gamma_5}$. The first non zero contribution is to be found at $n=4$, where there are finite and divergent contributions. The finite one leads to the following term,
\begin{equation}
	\mathcal A^{\slashed \partial \gamma_5}_{n=4,\mathrm{fin}} =  i(m^4 \mathcal I_i^4 - 4m^2 \mathcal I[q^2]_i^4 )\tr(\slashed D\slashed D\slashed D(\slashed \partial \theta) \gamma_5) \, ,
\end{equation}
using standard and convenient master integrals $\mathcal I$, written explicitly in Appendix~\ref{Appendix:master-integrals}. This contribution can be written as~\footnote{The trace over Dirac matrices is performed but the trace $\tr$ over the gauge group structure is left.},
\begin{equation}
	\mathcal A^{\slashed \partial \gamma_5}_{n=4,\mathrm{fin}} = -\dfrac{1}{32\pi^2}\tr\big(\slashed D\slashed D\slashed D(\slashed \partial \theta) \gamma_5\big) = \frac{i}{8\pi^2} \epsilon^{\mu\nu\rho\sigma}\tr(D_\mu D_\nu D_\rho (\partial_\sigma \theta)).
\end{equation}
The divergent contribution is regularised using BMHV scheme of dimensional regularisation~\cite{tHooft:1972tcz,Breitenlohner:1977hr}. With this choice, the divergent contribution is,
\begin{equation}
    \mathcal{A}^{\slashed \partial \gamma_5}_{n=4,\mathrm{div}} 
    = \dfrac{-i}{16\pi^2}\left(\varepsilon \frac{2}{\varepsilon}+ \mathcal O(\varepsilon)\right)\epsilon^{\mu\nu\rho\sigma}\tr(D_\mu D_\nu D_\rho (\partial_\sigma \theta)) \xrightarrow[\varepsilon \to 0]{} -\frac{i}{8\pi^2} \epsilon^{\mu\nu\rho\sigma}\tr(D_\mu D_\nu D_\rho (\partial_\sigma \theta)).
\end{equation}
The full $n=4$ contribution therefore cancels as the divergent and finite contributions compensate exactly,
\begin{equation}
    \mathcal{A}^{\slashed \partial \gamma_5}_{n=4} = \mathcal{A}^{\slashed \partial \gamma_5}_{n=4,\mathrm{div}}+ \mathcal{A}^{\slashed \partial \gamma_5}_{n=4,\mathrm{fin}} = 0 \;.
\end{equation}

Note that we talk about divergent contributions because the integrals are divergent, but in the end, the result is finite as the pole $2/\epsilon$ is compensated by an $\epsilon$ from the trace.

Turning now to the expansion of $\mathcal A^{m \gamma_5}$, its first contribution arises at $n=5$, and is fully finite. In this case there is no requirement to switch to $d$ dimensions~\footnote{for convenience with the notations, we are still using the master integrals which are technically defined in d-dimension.}. This contribution reads,
\begin{equation}
	\mathcal A^{m \gamma_5}_{n=5} = i\big( 2 m^6 \mathcal I_i^5  - 16 m^4 \mathcal I[q^2]_i^5 + 48 m^2 \mathcal I[q^4]_i^5 \big) \, \tr(\slashed D \slashed D \slashed D \slashed D \, \theta \gamma_5)
	\;.
\end{equation}
Performing the Dirac matrix algebra and using the expression of the master integrals given in appendix~\ref{Appendix:master-integrals}, the $n=5$ contributions reads,
\begin{equation}
	\mathcal A^{m \gamma_5}_{n=5} = -\dfrac{i}{16\pi^2}  \,\epsilon^{\mu\nu\rho\sigma} \theta \, \tr \, (F_{\mu\nu}F_{\rho\sigma})\, ,
\end{equation}
where the convention for the field strength is $F_{\mu\nu}=[D_\mu,D_\nu]$.

Within the CDE approach, this is the only surviving contribution, and it matches the well-known result for the axial current anomaly in a vector gauge field theory~\cite{Fujikawa:1979ay,Fujikawa:2004cx,Bardeen:1969md},
\begin{equation}
    \mathcal A = \mathcal A^{m \gamma_5} + \mathcal A^{\slashed \partial \gamma_5} = \mathcal A^{m \gamma_5}_{n=5} = -\dfrac{i}{8\pi^2} \theta \,  \tr \big( F_{\mu\nu} \tilde{F}^{\mu\nu} \big)
    \, , 
\label{ABJanomaly}
\end{equation}
where the convention for the dual tensor is $\tilde{F}^{\mu\nu} = 1/2\,\epsilon^{\mu\nu\rho\sigma}F_{\rho\sigma}$, with the choice $\epsilon^{0123} = +1$.

One may be a bit surprised by the fact that the anomaly ends up extracted from a non-divergent integral, for which no regularisation is needed. Let us stress though that the crucial step was to show that the $\mathcal A^{\slashed \partial \gamma_5}$ term gives no contribution in that particular case at order $m^0$.

Following a similar strategy, we will now discuss more generalities and details of the evaluation of the covariant and consistent anomalies in QFT based on a careful regularisation.

\subsection{ABJ anomaly in a given $2n$ dimensions from the CDE}
In this section, we provide a general approach to extend the computation of the ABJ anomaly in a given $2n$ dimensions. There is no obstruction to computing the anomaly for a given even dimension, but for arbitrary $2n$ dimensions it becomes more complicated to simplify Dirac traces. 

Starting from Eq.~\eqref{ano4d gamm5-first} in the manuscript, we have in $d=2n$ dimensions,
    \begin{align}
	\mathcal A =\int \frac{\dd^d q}{(2\pi)^d}\tr\, (2im\theta \gamma_{2n+1} + (\slashed \partial \theta) \gamma_{2n+1}) \sum_{k=0}^\infty  \tr \left[ \dfrac{-1}{\slashed q +m} \left(- i\slashed D  \right) \right]^k\dfrac{-1}{\slashed q +m}\ ,
	\label{ABJ: d=4}
    \end{align}
    where we have generalised the definition of $\gamma_5$ in $2n$ dimensions as,
    \begin{align}
            \gamma_{2n+1} = (i)^{n-1} \gamma^0\gamma^1\cdots \gamma^{2n-1}
            \, ,
    \end{align}
    and we have,
    \begin{align}
            \tr\bigg[ \gamma_{2n+1}\gamma^{\mu_1}\gamma^{\mu_2} \cdots \gamma^{\mu_{2n}} \bigg]
            &\equiv (-i)2^n \, \epsilon^{\mu_1\mu_2 \cdots \mu_{2n}}
            \, .
            \label{trace5: d=2n}
    \end{align}
    
By power counting, we can isolate the terms of order $m^0$,
    \begin{align}
    \mathcal A\bigg|_{d=2n} = \int \dfrac{\dd^{2n} q}{(2\pi)^{2n}}\tr\,\left[ \big(2im\theta \gamma_{2n+1} \big) 
    \bigg[\Delta \big(-i\slashed D\big) \bigg]^{2n} \Delta + \big[(\slashed\pa\theta)\gamma_{2n+1} \big] 
    \bigg[\Delta \big(-i\slashed D\big) \bigg]^{2n-1} \Delta \right]\, ,
    \label{ABJ-main: d=2n}
    \end{align}
    where $\Delta=-1/(\slashed q + m)$. 
    
We rewrite the propagators $\Delta$ in terms of bosonic and fermionic propagators $\Delta=\Delta_b+\Delta_f$ with $\Delta_b=m/(q^2-m^2)$ and $\Delta_f=-\slashed q/(q^2-m^2)$. The integration over momentum is non-vanishing for even powers in $q$, which means that we have to account for all the terms that have an even number of fermionic propagators. Therefore, the number of terms to compute increases significantly with the dimension.
    
For $2k$ fermionic propagators among the $2n+1$ propagators, we have to compute traces of the form,
    \begin{equation}
    g_{\alpha_1\dots\alpha_{2k}}\tr\,\gamma_{2n+1}\left( \gamma^{\alpha_1}\gamma^{\mu_1}\gamma^{\alpha_2}\gamma^{\mu_2}\dots\gamma^{\mu_{2n+1}}+\dots \right)\, ,
    \end{equation}
    where the dots encompass the remaining $\binom{2n+1}{2k}-1$ possible combinations of $2k$ fermionic propagators among $2n+1$ propagators, and $g_{\alpha_1\dots\alpha_{2k}}$ is the fully symmetrised metric\footnote{For example, $g_{\mu\nu\rho\sigma}=g_{\mu\nu}g_{\rho\sigma}+g_{\mu\rho}g_{\nu\sigma}+g_{\mu\sigma}g_{\nu\rho}$.}. Then such traces have to computed for all $k\leq n$.
    
Such a trace is not trivial to compute for arbitrary $k$ and $n$, which is why the general formula for the anomaly in $2n$ dimensions is not straightforward to obtain, and is out of the scope of this paper. For the computation in an arbitrary $2n$ dimensions, we refer the reader to Refs.~\cite{Banerjee:1985ti, Banerjee:1986bu}.

Within our framework, we can compute the ABJ anomaly (and other anomalies) in $4,\, 6,\, 8,\, \cdots$ dimensions, then extrapolate the result to $2n$ dimensions. This strategy is analogous to the computations of $l-$agon Feynman diagrams (with $l=n+1$) which have been performed by Frampton et al~\cite{Frampton:1983nr, Frampton:1983ah}.  
    
One must also generalise the definition of the master integrals in $2n$ dimensions, but this presents no difficulty.

\section{Anomalies in vector-axial gauge field theory}\label{sec:vectoraxial}\label{section3}

In the previous section, we have discussed the methodology to compute the Jacobian of a path integral measure by using EFT techniques, namely the CDE, and gave a concrete example by computing the well-known axial current anomaly in a vector gauge field theory. In this section, we apply this new formalism to recover the various and well-known anomalies in vector-axial gauge field theory. If $\theta$ is charged under the $SU(N)$ gauge group of the theory, then the anomaly can either be covariant (covariant anomaly), or respect the Wess-Zumino consistency conditions ~\cite{Wess:1971yu} (consistent anomaly). Our computations in the following sections are performed in Minkowski space-time~\footnote{The standard computations of Refs.~\cite{Bertlmann:1996xk, Fujikawa:2004cx} are performed in Euclidian space.}, and our results agree with the traditional ones (see for example Refs.~\cite{Bertlmann:1996xk, Fujikawa:2004cx, Bilal:2008qx}).

In our computation, it is necessary to consider $\theta$ local.  In practice, if $\theta\in SU(N)$ (and $V,A\in SU(N)$ as well) is associated to a global symmetry, we conduct the computation with $\theta$ local, but we should regularise in order to get the covariant anomaly. If $\theta\in SU(N)$ is associated to a local symmetry, i.e a gauge transformation, we should regularise in order to get the consistent anomaly (gauge anomaly).

\subsection{Definiteness and regularisation}
\label{Definiteness}

Consider the following Lagrangian,
\begin{equation}
    \mathcal{L}=\bar\psi(i\slashed\pa-\slashed V-\slashed A\gamma_5-m)\psi\, ,
\end{equation}
with $V_{\mu}$ and $A_{\mu}$ a vector and axial gauge field, elements of $SU(N)$~\footnote{In order to clarify our manuscript, we postpone the important discussion about manifest gauge or global symmetry invariance, the mass term as an hard breaking source in the unitary basis and the introduction of Goldstone bosons to implement spontaneous symmetry breaking to section~\ref{massterm}.}. 
It is anomalous under the fermion reparametrisation,
\begin{align}
\begin{split}
&\psi\to e^{i\theta(x)\gamma_5}\psi
\, , \quad
\bar\psi\to \bar\psi e^{i\theta(x)\gamma_5}\, ,
\label{FermReparam}
\end{split}
\end{align}
with $\theta$ infinitesimal, and can be charged under the $SU(N)$ gauge group.
The Jacobian of this reparametrisation can be expressed as follows,
\begin{equation}
J[\theta]=\frac{\det(i\sD-m)}{\det(i\sD-m-(\sD\theta)\gamma_5-2im\theta\gamma_5)}\, .
\label{JacGeneral}
\end{equation}
However, we know that the anomaly associated to the axial reparametrisation may as well appear in the vector current or the axial current (see for example Refs.~\cite{Weinberg:1996kr,Bertlmann:1996xk}). The Jacobian in Eq.~\eqref{JacGeneral} standing as it is can lead to any distribution of the anomaly in both currents.

Moreover, since the theory has an axial gauge field $A_{\mu}$, the reparametrisation in Eq.~\eqref{FermReparam} can be interpreted as a gauge transformation (i.e local transformation) if $\theta$ is charged under the gauge group.
For these reasons, the Jacobian in Eq.~\eqref{JacGeneral} is ill-defined.

To make sense of this ratio of formal determinants, we need to regularise it. In CDE, the most convenient regularisation scheme is dimensional regularisation. However, it is well-known that the definition of $\gamma_5$ in dimensional regularisation is ambiguous due to its intrinsic 4 dimensional nature~\cite{tHooft:1972tcz}.
We will propose two methods of regularising the Jacobian of 
Eq.~\eqref{JacGeneral}. The first method consists in working with the formal determinant in dimensional regularisation and, throughout the computation, deal with the ambiguity related to $\gamma_5$ using free parameters~\cite{Elias:1982ea,Quevillon:2021sfz}. The second method consists in bosonising the determinant, making it finite, hence fixing the ambiguity before the calculation. The first method can be seen as more general (or maybe naïve and brutal) as one first regularises an ill-defined quantity inserting as much freedom as needed and secondly call for coherence (covariance, integrability/consistence) of the obtained theory to fix those ambiguities. We believe that a remarkable advantage of this approach is that its derivation is smooth and self-consistent within the path integral formalism. The second method works the opposite way, as one firstly calls for a well defined theory (free of any ambiguity) and secondly perform the regularisation. As we will see both have their own advantages and disadvantages and we find it illuminating to present them both. We should also notice that while we believe the first method is novel in its approach, the bosonisation method is well-known~\cite{Banerjee:1986bu,Bertlmann:1996xk,Henning:2014wua}, however its combined used with the CDE to evaluate anomalies, is new and since this offers a powerful tool and interesting implications for EFTs related topics, it deserves to be duly studied here.

\subsubsection{Ambiguities and free parameters}
\label{sec: ambiguitites-free-parameters}

In $d$ dimensions, $\gamma_5$ is ill-defined. One cannot maintain both the cyclicity of the trace and Clifford algebra. There exist many ways of defining $\gamma_5$ in $d$ dimensions consistently~\cite{tHooft:1972tcz,Breitenlohner:1977hr,Chanowitz:1979zu,Novotny:1994yx,Elias:1982ea}, although they may yield different results. The ambiguity in the Jacobian of Eq.~\eqref{JacGeneral} lies in the dependence on the choice of the $\gamma_5$ regularisation scheme.

In a diagrammatic approach, the amplitude of a diagram is dictated by the Feynman rules. However, it does not specify by which vertex we should start writing the amplitude of the diagram, which results in different possible position for $\gamma_5$. Since in $d$ dimensions, the different positions of $\gamma_5$ are not equivalent, we have an ambiguity in the position of $\gamma_5$.

Nonetheless, it is possible to compute traces of $\gamma_5$ in $d$ dimensions while keeping track of the ambiguity by introducing free parameters~\cite{Elias:1982ea}. We outline the method in the following.

Consider the trace,
\begin{equation}
    \tr\,(\gamma_5\gamma^\mu\gamma^\nu\gamma^\rho\gamma^\sigma)\, .
\end{equation}
In 4 dimensions, one can use Clifford algebra to move the $\gamma_5$ at different positions,
\begin{equation}
     \tr\,(\gamma_5\gamma^\mu\gamma^\nu\gamma^\rho\gamma^\sigma)=\tr\,(\gamma^\mu\gamma^\nu\gamma_5\gamma^\rho\gamma^\sigma)=\tr\,(\gamma^\mu\gamma^\nu\gamma^\rho\gamma^\sigma\gamma_5)\, .
\end{equation}
However, this may not be true anymore in $d$ dimensions. 
For example if we use BMHV scheme~\cite{Breitenlohner:1977hr}, we maintain the cyclicity of the trace but we have to abandon Clifford algebra. We then have an ambiguity on the position of $\gamma_5$ in the trace. The trick presented in Ref.~\cite{Elias:1982ea} consists in implementing all the positions for $\gamma_5$ that are equivalent in 4 dimensions, with a free parameter for each,
\begin{equation}
\tr\,(\gamma_5\gamma^\mu\gamma^\nu\gamma^\rho\gamma^\sigma)\to\alpha\,\tr\,(\gamma_5\gamma^\mu\gamma^\nu\gamma^\rho\gamma^\sigma)+\beta\,\tr\,(\gamma^\mu\gamma^\nu\gamma_5\gamma^\rho\gamma^\sigma)+\delta\,\tr\,(\gamma^\mu\gamma^\nu\gamma^\rho\gamma^\sigma\gamma_5)\, ,  
\end{equation}
with the condition $\alpha+\beta+\delta=1$, so that we recover $\tr\,(\gamma_5\gamma^\mu\gamma^\nu\gamma^\rho\gamma^\sigma)$ in 4 dimensions.

The introduction of those free parameters with all the equivalent positions (in 4 dimensions) of $\gamma_5$ makes the trace regularisation scheme independent. Therefore, we can choose a specific scheme to compute each separate trace. If the result depends on the free parameters in the end, it means that the initial trace itself is ambiguous.

For the example above, we compute each separate trace using BMHV scheme to get,
\begin{align}
\tr\,(\gamma_5\gamma^\mu\gamma^\nu\gamma^\rho\gamma^\sigma) \rightarrow & \,\alpha\,\tr\,(\gamma_5\gamma^\mu\gamma^\nu\gamma^\rho\gamma^\sigma)+\beta\,\tr\,(\gamma^\mu\gamma^\nu\gamma_5\gamma^\rho\gamma^\sigma)+\delta\,\tr\,(\gamma^\mu\gamma^\nu\gamma^\rho\gamma^\sigma\gamma_5)
\nonumber \\
=&(\alpha+\beta+\delta)(-4i\epsilon^{\mu\nu\rho\sigma})
=-4i\epsilon^{\mu\nu\rho\sigma}\, ,
\end{align}
where we have used the condition $\alpha+\beta+\delta=1$ to match with the result in 4 dimensions. It turns out that this trace is non-ambiguous.

However, consider the following trace with one contraction among the Dirac matrices,
\begin{align}
&\tr\,(\gamma_5\gamma^\mu\gamma^\nu\gamma^a\gamma^\rho\gamma^\sigma\gamma_a)\nonumber\\
\to&\,\alpha\tr\,(\gamma_5\gamma^\mu\gamma^\nu\gamma^a\gamma^\rho\gamma^\sigma\gamma_a)+\beta\tr\,(\gamma^\mu\gamma^\nu\gamma_5\gamma^a\gamma^\rho\gamma^\sigma\gamma_a)+\gamma\tr\,(\gamma^\mu\gamma^\nu\gamma^a\gamma^\rho\gamma_5\gamma^\sigma\gamma_a)+\delta\tr\,(\gamma^\mu\gamma^\nu\gamma^a\gamma^\rho\gamma^\sigma\gamma_a\gamma_5)
\nonumber\\
=&(-1+2\gamma) \, 4i(d-4)\epsilon^{\mu\nu\rho\sigma}\, .
\end{align}
It is ambiguous because even after enforcing the condition $\alpha+\beta+\gamma+\delta=1$, the result still depends on a free parameter. Actually, insofar as there is more than one contraction among the Dirac matrices, the trace will be ambiguous\footnote{See appendix \ref{appfreeparam} for the case with two contractions among the sequence of Dirac matrices.}.
As a consequence, when computing the anomaly, the final result depends on free parameters. Those free parameters are then fixed under physical constraints, for example by enforcing gauge invariance and vector current conservation.

Although the positions of $\gamma_5$ in the computation of the path integral Jacobian are not arbitrary, as opposed to the diagrammatic approach, it may still bear traces that depends on the choice of $\gamma_5$ scheme. Despite the absence of arbitrariness in the position of $\gamma_5$ we will still rely on the free parameters trick to compute the ambiguous Jacobian, since it allows us to compute the traces in a $\gamma_5$ scheme independent way.

\subsubsection{A well-known treatment : the bosonisation}
\label{DetEigReg}

Before delving into the expansion of the determinant, it is possible to regularise it. One way of achieving a regularised Jacobian is to bosonise it.

\paragraph{Vector gauge theory:}

Consider first a vector gauge theory, the Jacobian can be squared to bosonise it,
\begin{equation}
\mathcal{L}=\bar\psi(i\slashed\pa-\slashed V-m)\psi\, .
\end{equation}
We will show in section \ref{consano} that the Jacobian in Eq.~\eqref{JacGeneral} can be written as,
\begin{equation}
J[\theta]^2=\frac{\det(\sD^2+m^2)}{\det(\sD^2+m^2+\{i\sD,(\sD\theta)\gamma_5\}+4im^2\theta\gamma_5)}\, .
\label{VectorlikeBosonized}
\end{equation}
This Jacobian yields the same result as the fermionic Jacobian Eq.~\eqref{Jacdet} insofar as the theory is not chiral.

\paragraph{Vector-axial theory:}
Now, consider a vector-axial gauge theory,
\begin{equation}
    i\sD=i\slashed\pa-\slashed V-\slashed A\gamma_5\, .
\end{equation}

Now the operator $i\sD-m$ does not have a well-defined eigenvalue problem, the presence of the axial field spoils the hermitianity. It is however crucial to have a well-defined eigenvalue problem to make sense of the determinant, which is the product of the eigenvalues of the operator. 

We will now present a solution for bosonising the Jacobian of Eq.~\eqref{JacGeneral} that let us deal with hermitian and gauge covariant operators.

One way to obtain a hermitian operator is to use the following Laplace operators,
\begin{equation}
\sD^\dagger\sD \,\text{ and }\,\sD\sD^\dagger\, .
\end{equation}
These operators are hermitian, hence have a well-defined eigenvalue problem. They preserve the spectrum of the theory (see for example Ref.~\cite{Bertlmann:1996xk}), hence do not change the value of the determinants (aside squaring them). Besides, they lead to a gauge covariant regularisation of the bosonised form of the Jacobian.

We will show in section \ref{sectioncovanomaly}, that the Jacobian of Eq.~\eqref{JacGeneral} can be written as,
\begin{equation}
J[\theta]^2=\frac{\det \left(-(i\sD)^\dagger i\sD+m^2\right)}{\det \left(-(i\sD)^\dagger i\sD+m^2+f(\theta)\right)}\, ,
\end{equation}
where,
\begin{equation}
f(\theta)=4im^2\theta\gamma_5-i[\theta,-D^2]\gamma_5-\half[\sigma.F^V,\theta]\gamma_5-\half[\sigma.F^A\gamma_5,\theta]\gamma_5 \ .
\end{equation}
$F^V$ and $F^A$ are the Bardeen curvatures defined a bit later in Eqs.~\eqref{eq:FV} and \eqref{eq:FA}.  

This bosonised determinant is finite hence unambiguous. Besides, since the regularisation it provides is gauge covariant~\cite{Bertlmann:1996xk}, the final result can only be gauge covariant, hence the so-called covariant anomaly. 

On the other hand, if we want to compute the consistent anomaly, we can try to use the bosonisation as in the vector gauge theory. However, the operator $\sD^2+m^2$ is still not hermitian. We palliate this problem using the analytic continuation $A_\mu\to iA_\mu$ that restores the hermitianity of $i\sD$, hence of $\sD^2+m^2$.

The Jacobian will then be written as,
\begin{equation}
J[\theta]^2=\frac{\det(\sD^2+m^2)}{\det(\sD^2+m^2+\{i\sD,(\sD\theta)\gamma_5\}+4im^2\theta\gamma_5)}\, .
\end{equation}
Unfortunately, as we will see, this does not suffice to fix the ambiguity. It does not necessarily yield the consistent anomaly.

\subsection{A generic Lagrangian}

To pave the way for our computations of covariant and consistent anomalies, we present briefly the generic Lagrangian we will consider and our notations.
One can consider a gauge theory in which left and right-handed fermion components are charged under a non-Abelian gauge group, then described by the following Lagrangian,
\begin{equation}
    \mathcal L = \bar \psi_L \gamma^{\mu}(i\partial_{\mu} - L_{\mu}) \psi_L + \bar \psi_R \gamma^{\mu}(i\partial_{\mu} - R_{\mu}) \psi_R -m \bar \psi \psi,
\end{equation}
where $L_\mu = L_{\mu}^aT^a$ and $R_\mu = R_{\mu}^aT^a$ are gauge fields belonging to $SU(N)$. In term of the projector algebra, this Lagrangian can be written in terms of vector-axial gauge fields as follows,
\begin{equation}\label{eq:toylagrangian}
    \mathcal L = \bar \psi (i\slashed \partial - \slashed L P_L - \slashed R P_R - m)\psi = \bar \psi (i\slashed \partial - \slashed V - \slashed A \gamma_5 -m)\psi \equiv \bar \psi (i \slashed D -m)\psi ,
\end{equation}
where we defined the fields $V_\mu$, $A_\mu$, and the covariant derivative as follows,
\begin{equation}
    V_\mu \equiv \frac{L_\mu + R_\mu}{2}
    \, , \quad 
    A_\mu \equiv \frac{R_\mu - L_\mu}{2}
    \, , \quad
    iD_\mu \equiv i\big( \partial_\mu + iV_\mu + iA_\mu\gamma_5 \big)
    \, .
    \label{definitions: V,A,CovD}
\end{equation}
The computation of the commutator $[D_\mu,D_\nu]$ permits to define two Bardeen's curvatures (see Ref.~\cite{Bardeen:1969md}) by identifying the axial and vector part such that $[D_\mu,D_\nu]\equiv F^V_{\mu\nu} + F^A_{\mu\nu}\gamma_5$, which leads to the following expressions,
\begin{equation}\label{eq:FV}
    F^V_{\mu\nu} = i\left((\partial_\mu V_\nu) - (\partial_\nu V_\mu) + i [V_\mu,V_\nu] +i[A_\mu,A_\nu]\right),
\end{equation}
\begin{equation}\label{eq:FA}
    F^A_{\mu\nu} = i\left((\partial_\mu A_\nu) - (\partial_\nu A_\mu) + i [A_\mu,V_\nu] +i[V_\mu,A_\nu]\right).
\end{equation}
In the L/R basis the field strengths are,
\begin{equation}\label{eq:FL}
    F^{L}_{\mu\nu}=i\left((\partial_\mu L_\nu)-(\partial_\mu L_\mu) +i[L_\mu,L_\nu]\right)
\end{equation}
\begin{equation}\label{eq:FR}
    F^{R}_{\mu\nu}=i\left((\partial_\mu R_\nu)-(\partial_\mu R_\mu) +i[R_\mu,R_\nu]\right) \ ,
\end{equation}
and the Bardeen curvatures are related to the L/R curvatures by,
\begin{equation}\label{eq:VtoLR}
F^V_{\mu\nu}=\half (F^R_{\mu\nu}+F^L_{\mu\nu})
\end{equation}
\begin{equation}\label{eq:AtoLR}
F^A_{\mu\nu}=\half (F^R_{\mu\nu}-F^L_{\mu\nu}) \ .
\end{equation}

\subsection{Covariant anomaly} \label{sectioncovanomaly}

\subsubsection{Mass term, manifest symmetry invariance and Goldstone bosons} \label{massterm}

All along our work, we constantly integrate out a massive chiral fermion. The mass term is a hard breaking source of axial symmetries (local or global). In order to make manifest those symmetries at tree-level one can evidently implement their spontaneous breaking introducing then their associated Goldstone bosons. We chose to work within the unitary basis and loose manifest tree-level axial invariance (when relevant) in order to deal with simpler functional determinants. The Goldstone bosons will be explicitly re-introduced only when it is necessary, see section~\ref{subsec:vector-global-anomalous}. Consequently, one should not be surprised if we discuss an anomalous global symmetry which looks naively already broken at tree-level.~\footnote{A detailed discussion on the parametrisation of local and global anomalous symmetries can be find in Ref.~\cite{Quevillon:2021sfz}.}

\subsubsection{Case of an anomalous axial symmetry}\label{subsec:axial-global-anomalous}
We state here again, for convenience, the Lagrangian and the Jacobian associated to the axial transformation.

Starting from the vector-axial Lagrangian of Eq.~\eqref{eq:toylagrangian}, let us perform an axial fermion reparametrisation,
\begin{equation}
	\psi \to e^{i\theta(x)\gamma_5} \psi, \quad \bar\psi \to \bar \psi\, e^{i\theta(x) \gamma_5} \ .
\end{equation}
Under this fermion reparametrisation, the Lagrangian given by Eq.~\eqref{eq:toylagrangian} becomes,
\begin{equation}
	\mathcal L \to \bar \psi \big[ i\slashed D - m - 2im\,\theta(x)\gamma_5 - \big(\slashed D\theta\gamma_5\big) \big] \psi 	\ ,
\end{equation}
where the quantity inside the parenthesis, $\big(\slashed D\theta\gamma_5\big) = \left((\slashed\pa \theta)+i[\slashed V,\theta]+i[\slashed A,\theta]\gamma_5\right)\gamma_5$, indicates that the covariant derivative locally acts on $\theta(x)$ (i.e not on everything on its right). The Jacobian produced by this transformation is therefore given by the following expression,
\begin{equation}
	J[\theta] = \dfrac{\det \big(i \slashed D -m\big)}{\det e^{i\theta(x)\gamma_5} \big(i\slashed D -m\big) e^{i\theta(x)\gamma_5}} = \dfrac{\det \big(i \slashed D -m\big)}{\det\big(i\slashed D -m - 2im\theta \gamma_5 - (\slashed D\theta\gamma_5)\big)} \ .
	\label{Jacobian: axial-global-rotation}
\end{equation}
As emphasised in the previous sections, this Jacobian is ill-defined. The next step is to explicitly compute it, according to the methods proposed in section \ref{Definiteness}.\\

\paragraph{Fermionic expansion with free parameters}
\label{FermExpFreeParam}

We are now in the situation where we are looking to evaluate an equivalent of Eq.~\eqref{ano4d} for a vector and axial gauge field theory,
\begin{equation}
    \mathcal{A}=\int\frac{\dd^dq}{(2\pi)^d}\tr\,\left((\slashed D\theta)\gamma_5+2im\theta\gamma_5\right)\sum_{n\geq0}\left[\frac{-1}{\slashed q-m}(-i\sD)\right]^n\frac{-1}{\slashed q-m
}\, ,
\end{equation}
where $\theta$ belongs to $SU(N)$ and the covariant derivative is $D_\mu=\pa_\mu+iV_\mu+iA_\mu\gamma_5$.
\newline
\newline
Let's start by computing the mass term. The integrals are finite hence no ambiguity arises from this term.

The propagators that appear in the expansion need to be expanded as $-1/(\slashed q+m)=\Delta_b+\Delta_f$ where the bosonic propagator is $\Delta_b=m/(q^2-m^2)$ and the fermionic propagator is $\Delta_f=-\slashed q/(q^2-m^2)$. The integrals over momentum are non-vanishing only if the integrand has an even power in $q$ in the numerator (the denominator always has an even power in $q$). Therefore, the number of fermionic propagators must be even. This leaves us with three contributions. Note that each of those contributions is finite, thus the computation is performed in 4 dimensions. 

$\bullet$ The contribution to the anomalous interaction involving only bosonic propagators is,
\begin{equation}
m^5\mathcal{I}[q^0]^52im \tr\left(\theta\gamma_5\sD^4\right)\, .
\end{equation}

$\bullet$ The contribution to the anomalous interaction involving  two fermionic propagators is,
\begin{equation}
m^3\mathcal{I}[q^2]^52im \tr\left(\theta\gamma_5[\gamma^a\sD\gamma_a\sD^3+\gamma^a\sD^2\gamma_a\sD^2+...]\right)\, ,
\end{equation}
where the dots bear all the remaining insertions of the two fermionic propagators ($\binom{5}{2}=10$ combinations).

$\bullet$ The contribution to the anomalous interaction involving four fermionic propagators is,
\begin{equation}
m\mathcal{I}[q^4]^52im \tr\left(\theta g_{abcd}\gamma_5[\gamma^a\sD\gamma^b\sD\gamma^c\sD\gamma^d\sD+\gamma^a\sD\gamma^b\sD\gamma^c\sD\sD\gamma^d+...]\right)\, ,
\end{equation}
where again the dots bear all the remaining insertions of the four fermionic propagators ($\binom{5}{4}=5$ combinations) and $g_{abcd}=g_{ab}g_{cd}+g_{ac}g_{bd}+g_{ad}g_{bc}$.

Then one needs to expand the covariant derivatives in order to extract the $\gamma_5$ from the axial fields and compute the Dirac traces. It is then simple algebra to form the field strengths as defined in Eqs.~\eqref{eq:FV} and \eqref{eq:FA}.

The mass term then yields a contribution that corresponds to the so-called Bardeen anomaly (with conserved vector current), that is to say the consistent anomaly,
\begin{align}
\begin{split}
\mathcal{A}^{m\gamma_5}=\frac{-i}{16\pi^2}\epsilon^{\mu\nu\rho\sigma}\tr\, \theta^aT^a\bigg(& F^V_{\mu\nu}F^V_{\rho\sigma}+\frac{1}{3}F^A_{\mu\nu}F^A_{\rho\sigma}\\
&-\frac{8}{3}\left( iA_\mu iA_\nu F^V_{\rho\sigma} +iA_\mu F^V_{\nu\rho} iA_\sigma +F^V_{\mu\nu}iA_\rho iA_\sigma \right) +\frac{32}{3} iA_\mu iA_\nu iA_\rho iA_\sigma \bigg)\\
&\hspace{-3.7cm}=\mathcal{A^{\mathrm{Bardeen}}}\, .
\label{BardeenAn}
\end{split}
\end{align}
Now let's focus on the derivative term,
\begin{equation}
\mathcal{A}_{\slashed\pa\gamma_5}=\int\frac{\dd^dq}{(2\pi)^d}\tr\,(\slashed D\theta)\gamma_5\sum_{n\geq0}\left[\frac{-1}{\slashed q-m}(-i\sD)\right]^n\frac{-1}{\slashed q-m
}\, .
\end{equation}
We proceed similarly for the derivative term to obtain the following contributions:
\begin{itemize}
    \item The contribution to the anomalous interaction involving only bosonic propagators is,
\begin{equation}
im^4\mathcal{I}[q^0]^4\tr\,\left((\sD\theta)\gamma_5(\sD)^3\right)\, .
\label{CovADerivativeTermBos}
\end{equation}
    
    \item The contribution to the anomalous interaction involving two fermionic propagators is,
\begin{equation}
im^2\mathcal{I}[q^2]^4\tr\,\left( (\sD\theta)\gamma_5 [\gamma^a\sD\gamma_a\sD\sD+\gamma^a\sD\sD\gamma_a\sD+\dots] \right)\, ,
\label{CovADerivativeTerm2Ferm}
\end{equation}
where the dots denote the other $\binom{4}{2}=6$ combinations for the insertions of the two fermionic propagators.

    \item The contribution to the anomalous interaction involving  four fermionic propagators is,
\begin{equation}
i\mathcal{I}[q^4]^4\tr\,\left( (\sD\theta)\gamma_5 [\gamma^a\sD\gamma^b\sD\gamma^c\sD\gamma^d g_{abcd}] \right)\, .
\label{CovADerivativeTerm4Ferm}
\end{equation}
\end{itemize}
Now this last integral is divergent, thus the trace that appear in the term with four fermionic propagators is ambiguous. We use the trick described in section \ref{sec: ambiguitites-free-parameters} to keep track of the ambiguity. Therefore, the three contributions above may be written, after integrating by parts, as a sum of operators with a free parameter for each. The result can thus be written fully in terms of free parameters associated to each possible operator (the finite contributions will just combine with a free parameter to give a different free parameter). We thus have,
\begin{equation}
\mathcal{A}_{\slashed\pa\gamma_5}=\frac{-i}{16\pi^2}\epsilon^{\mu\nu\rho\sigma}\tr\,\theta^a T^a\left(\sum_i a_i X_{i,\,\mu\nu\rho\sigma} \right)\, ,
\end{equation}
where $X_i$ are all the possible operators of the form $\mathcal{O}_1\mathcal{O}_2\mathcal{O}_3\mathcal{O}_4$ with $\mathcal{O}_{1\leq i\leq4}\in\{V,A,\pa\}$ that can be formed, provided it has an even number of $A$ fields (the number of $\gamma_5$ must be odd). Note that the operators with a partial derivative to the right vanish, and those with consecutive partial derivatives vanish due to the contraction with the $\epsilon$ tensor. This leaves us with 22 possible operators, with 22 free parameters $a_i$. 
\newline
\newline
We then want to enforce the covariance of $\mathcal{A}_{m\gamma_5}+\mathcal{A}_{\slashed\pa\gamma_5}$ under the gauge transformation,
\begin{align}
\begin{cases}
    V_{\mu} \rightarrow V_{\mu} + (D^V_{\mu}\varepsilon_V) + i[ A_{\mu},\varepsilon_A ]
    \\ 
    A_{\mu} \rightarrow A_{\mu} + i[A_{\mu},\varepsilon_V]+ (D^V_{\mu}\varepsilon_A) 
\end{cases} \, .
\label{gaugevarVA}
\end{align}
We can focus only on the gauge transformation associated to $\epsilon_A$, it will be sufficient to fix the free parameters.

A covariant operator $\mathcal{O}$ must transform as,
\begin{equation}
\delta\mathcal{O}=[\epsilon_A,\mathcal{O}]\, ,
\label{GaugeCov}
\end{equation}
under the $\epsilon_A$ gauge transformation. Therefore, we enforce that the terms with derivatives of $\epsilon_A$ vanish, and also that $\epsilon_A$ must appear either at the beginning or at the end of each operator.

For example, after performing the gauge variation we have, among others, the following operator,
\begin{equation}
f(a_1,\dots,a_{22}) \epsilon^{\mu\nu\rho\sigma} (\pa_\mu A_\nu) \epsilon_A V_\rho V_\sigma\, ,
\end{equation}
where $f$ is some linear function of the free parameters.
This term must vanish for the result to be gauge covariant because $\epsilon_A$ is sandwished between operators, hence it cannot occur from a term of the form Eq.~\eqref{GaugeCov}. We hence obtain a constraint on the free parameters.

It turns out that enforcing these conditions fixes 21 free parameters out of 22. We rely on the result from the ABJ anomaly to fix the last free parameter that we call $\beta$. Setting $A=0$ we are left with,
\begin{equation}
\frac{-i}{16\pi^2}(1+\beta)\epsilon^{\mu\nu\rho\sigma}F_{\mu\nu}F_{\rho\sigma}\, ,
\label{comptoABJ}
\end{equation}
where $F_{\mu\nu}=(\pa_\mu i V_\nu)-(\pa_\nu i V_\mu)+[iV_\mu,iV_\nu]$. Eq.~\eqref{comptoABJ} is covariant regardless of the normalisation, this is why it needs to be compared with the anomaly in a vector-like theory to fix $\beta$ (i.e the ABJ anomaly). This amounts to enforcing the conservation of the vector current. We thus deduce that $\beta=0$. Now that all the free parameters are fixed, we obtain,
\begin{equation}
\mathcal{A}_{m\gamma_5}+\mathcal{A}_{\slashed\pa\gamma_5}=\frac{-i}{16\pi^2}\epsilon^{\mu\nu\rho\sigma}\tr\,\theta^a T^a\left( F^V_{\mu\nu}F^V_{\rho\sigma}+F^A_{\mu\nu}F^A_{\rho\sigma} \right)\, ,
\end{equation}
with $F^V$ and $F^A$ the Bardeen curvatures as defined in Eqs.~\eqref{eq:FV},\eqref{eq:FA}. This is the covariant non-Abelian anomaly in the axial current in a vector-axial theory.
Note that the relative coefficient between $F^V\tilde F^V$ and $F^A\tilde F^A$ is fixed by requiring the covariance of the result, since $F^V\tilde F^V+b F^A\tilde F^A$ is not covariant unless $b=1$. 

It can also be written in the L-R basis as,
\begin{equation}
\mathcal{A}_{m\gamma_5}+\mathcal{A}_{\slashed\pa\gamma_5}=\frac{-i}{32\pi^2}\epsilon^{\mu\nu\rho\sigma}\tr\,\theta^a T^a\left( F^L_{\mu\nu}F^L_{\rho\sigma}+F^R_{\mu\nu}F^R_{\rho\sigma} \right)\, .   
\end{equation}

Finally, we can mention the Bardeen-Zumino polynomial (BZ polynomial) \cite{Bardeen:1984pm} that naturally appears in our computation. The BZ polynomial is the unique local function $\mathcal{P}^\mu$ such that the gauge variation of $D_\mu\mathcal{P}^\mu$ cancels exactly the gauge variation of the consistent anomaly.

The ambiguity in the derivative term $\mathcal{A}_{\slashed\pa \gamma_5 }$ was fixed by requiring that the mass term and derivative term together are gauge covariant, that is to say, that the gauge variation of the derivative term cancels exactly the gauge variation of the unambiguous mass term. Since the mass term coincides with the consistent anomaly, then the derivative term cancelling its gauge variation is by definition the divergence of the BZ polynomial. 

The derivative term thus reads,
\begin{align}
\begin{split}
\mathcal{A}_{\slashed\pa\gamma_5}&=\theta^a (D_\mu \mathcal{P}^\mu)^a\\
&=\frac{-i}{16\pi^2}\epsilon^{\mu\nu\rho\sigma}\tr\,\theta\left( \frac{2}{3}F^A_{\mu\nu} F^A_{\rho\sigma}+\frac{8}{3}\left( A_\mu A_\nu F^V_{\rho\sigma} + A_\mu F^V_{\nu\rho}A_\sigma + F^V_{\mu\nu}A_\rho A_\sigma \right)-\frac{32}{3}A_\mu A_\nu A_\rho A_\sigma \right)\, .
\end{split}
\end{align}

Note that the divergence of the BZ polynomial was not obtained by substracting the covariant anomaly to the consistent anomaly, but truly by requiring the cancelation of the gauge variation of the consistent anomaly.

Eventually, we emphasise that the BZ polynomial itself can be obtained by summing Eqs.~\eqref{CovADerivativeTermBos}, \eqref{CovADerivativeTerm2Ferm} and \eqref{CovADerivativeTerm4Ferm}, performing the Dirac trace using the values of the free parameters that cancel the gauge variation of the anomaly, as we did. We thus obtain a term of the form $(D_\mu\theta)\mathcal{P}^\mu$ where $\mathcal{P}^\mu$ is the BZ polynomial.

\paragraph{Bosonisation method.} \label{covbosform}

The previous method proceeds by carrying dimensional regularisation on the ill-defined functional determinants of the Jacobian of Eq.~\eqref{Jacobian: axial-global-rotation}. A main difference with Fujikawa's approach is that one does not need to directly worry about whether the Dirac operator has a well defined eigenvalue problem, and then compute its spectrum. However, it exists a known trick which consists in transforming that Jacobian into a another well suited quantity, a Jacobian ``squared''. 

As suggested in Refs.~\cite{ Banerjee:1986bu,Banerjee:1990tv,Banerjee:1985ti,Ball:1988xg}, the operator $\slashed D^\dagger \slashed D$ and $\slashed D \slashed D^\dagger$ define a good eigenvalue problem in order to compute the spectrum of $i\slashed D$. In particular, since those two operators are Hermitian and covariant, they admit two orthogonal eigenbasis with real eigenvalues. For simplicity, we introduce $P_\mu=i D_\mu$, we then have,
\begin{equation}
	\sP^\dagger\sP \phi_n = \lambda_n^2 \phi_n,\quad \sP\sP^\dagger\varphi_n= \lambda_n^2 \varphi_n, \quad n\in \mathbb{N}, \quad \lambda_n\in \mathbb{R}.
\end{equation}
where
\begin{equation}
\sP\phi_n=\lambda_n\varphi_n\,\text{, }\,\sP^\dagger\varphi_n=\lambda_n\phi_n\,\text{ with }\,\lambda_n\in\mathds{R} \ .
\end{equation}
In order to form such operators from the orginal Jacobian Eq.~\eqref{JacGeneral}, one can build the following quantity,
\begin{align}
\begin{split}
J^2[\theta] &=\frac{\det (\sP^\dagger-m)}{\det\left( e^{i\theta\gamma_5}(\sP^\dagger-m)e^{i\theta\gamma_5}\right)}\frac{\det (\sP-m)}{\det\left( e^{i\theta\gamma_5}(\sP-m)e^{i\theta\gamma_5}\right)}\\
&=\frac{\det (\sP^\dagger-m)}{\det\left( e^{i\theta\gamma_5}(\sP^\dagger-m)e^{i\theta\gamma_5}\right)}\frac{\det (-\sP-m)}{\det \left(e^{i\theta\gamma_5}(-\sP-m)e^{i\theta\gamma_5}\right)}\\
&=\frac{\det \left(-\sP^\dagger \sP+m^2\right)}{\det \left(-\sP^\dagger \sP+m^2+m(\sP-\sP^\dagger)+f(\theta)\right)} \ ,
\label{jacboson}
\end{split}
\end{align}
where in the second line we have used the invariance of the determinant under the change $\gamma^\mu\to -\gamma^\mu$~\footnote{\label{footnoteSignFlip} It uses the fact that $\det=\Tr\exp$, and that the trace of an odd number of Dirac matrices always vanishes. Since they have to come in by pairs, the sign flip does not affect the result. Under this sign flip, $\gamma_5$ is unchanged since it has an even number of Dirac matrices.}, and we have defined,
\begin{align}
\begin{split}
f(\theta)=&4im^2\theta\gamma_5-i[\theta,P^2]\gamma_5-\half[\sigma.F^V,\theta]\gamma_5-\half[\sigma.F^A\gamma_5,\theta]\gamma_5\\
&+2im\left(\theta\gamma_5\sP-\sP\theta\gamma_5\right)+im\left((\sP\theta)-(\sP^\dagger\theta))\right)\gamma_5 \ .
\end{split}
\end{align}
$F^V$ and $F^A$ are the Bardeen curvatures as defined in Eq.~\eqref{eq:FV},~\eqref{eq:FA}, and $\sigma^{\mu\nu}=\frac{i}{2}[\gamma^\mu,\gamma^\nu]$. Details of the bosonisation are provided in appendix \ref{appbosonized}. 

$\theta$ is charged under the gauge group so $(\sP\theta)=(i\slashed\pa\theta)-[\slashed V,\theta]-[\slashed A\gamma_5,\theta]$. Therefore, we can a priori obtain the consistent or the covariant anomaly,  but we will see that the bosonisation we have chosen selects the covariant result.

We shamelessly used the multiplicativity property, $\det(A)\det(B)=\det(AB)$,  on non-regularised determinants. Indeed, this has to be admitted since CDE assumes that for a non-regularised determinant one can write $\log\det=\Tr\log$.\\

The computation of this Jacobian can be performed following the same principle given in section~\ref{sec:Main},
\begin{equation}
	2 \mathcal A = -\int \frac{\dd^d q}{(2\pi)^d} e^{iqx} \tr \left(f(\theta)+m(\sP-\sP^\dagger)\right)\frac{1}{-\sP^\dagger \sP+m^2} e^{-iqx} \ .
\end{equation}
At this point, one can recall that the terms that have an odd number of Dirac matrices vanish under the trace. Therefore, in the above we can drop the term $m(\sP-\sP^\dagger)$, and the terms $2im\left(\theta\gamma_5\sP-\sP\theta\gamma_5\right)$ and $im\left((\sP\theta)-(\sP^\dagger\theta)\right)\gamma_5$ from $f(\theta)$, since $1/(-\sP^\dagger\sP+m^2)$ has en even number of Dirac matrices. We are then left with,
\begin{equation}
	2 \mathcal A = -\int \frac{\dd^d q}{(2\pi)^d} e^{-iqx} \tr \left(-i[\theta,P^2]\gamma_5-\half[\sigma.F^V,\theta]\gamma_5-\half[\sigma.F^A\gamma_5,\theta]\gamma_5+4im^2\theta\gamma_5\right)\frac{1}{-\sP^\dagger\sP+m^2} e^{iqx}.
\end{equation}
This produces in the end,
\begin{equation}
2 \mathcal{A} = \int \frac{\dd^d q}{(2\pi)^d} \tr\, h(\theta)\sum_{n\geq0} \left[\Delta\left(-P^2+\frac{i}{2}\sigma.F^V+\frac{i}{2}\sigma.F^A\gamma_5+2q\cdot P\right)\right]^n\Delta,
\end{equation}
where
\begin{equation}
h(\theta)=-i[\theta,P^2]\gamma_5-\half[\sigma.F^V,\theta]\gamma_5-\half[\sigma.F^A\gamma_5,\theta]\gamma_5+4im^2\theta\gamma_5 \ ,
\end{equation}
and $\Delta=1/(q^2-m^2)$.

After scrutinizing the various terms, they appear to be all finite (hence non-ambiguous), and in the end only one term contributes,
\begin{align}
2\mathcal{A}&=\int\frac{\dd^d q}{(2\pi)^d}\Delta^3\tr\,4im^2\theta\gamma_5(\frac{i}{2}\sigma.F^V+\frac{i}{2}\sigma.F^A\gamma_5)^2 \nonumber \\
&=2\frac{-i}{16\pi^2}\epsilon^{\mu\nu\rho\sigma}\tr\,\theta(F^V_{\mu\nu}F^V_{\rho\sigma}+F^A_{\mu\nu}F^A_{\rho\sigma})\, ,
\end{align}
where we have discarded terms with even number of $\gamma_5$ matrices (they cannot yield a boundary term so cannot contribute to the final result). We thus obtain the so-called covariant anomaly,
\begin{align}
\begin{split}
\mathcal{A}&=\frac{-i}{16\pi^2}\epsilon^{\mu\nu\rho\sigma} \tr \, \theta (F^V_{\mu\nu}F^V_{\rho\sigma}+F^A_{\mu\nu}F^A_{\rho\sigma})
= \frac{-i}{32\pi^2}\epsilon^{\mu\nu\rho\sigma} \tr \, \theta (F^L_{\mu\nu}F^L_{\rho\sigma}+F^R_{\mu\nu}F^R_{\rho\sigma}) \ .
\end{split}
\label{covabos}
\end{align}
Additional details on the calculation are provided in the appendix~\ref{appbosonized}.\\

As an important remark, in the bosonised form of the Jacobian, it turns out that the derivative coupling contribution vanishes at order $m^0$, only the mass term $4im^2\theta\gamma_5$ (which stems from the term $2im\theta\gamma_5$ before bosonising) contributes.
As a result, the computation is finite (in the sense that no divergent integral appears), therefore no ambiguity arises due to the definition of $\gamma_5$ since we can perform the calculation in 4 dimensions. Although the operators stemming from the derivative coupling ($(\slashed D\theta)\gamma_5$ before bosonising) do not contribute to the anomaly, they are required to compensate the finite higher order (of order $1/m^k$ with $k>0$) terms in the mass expansion, since the final result has to be exact at order $m^0$. Note also that we bosonised using $\sD^\dagger\sD$, but we could have equivalently used $\sD\sD^\dagger$ to get the same result.

\subsubsection{Case of an anomalous vector symmetry}\label{subsec:vector-global-anomalous}

Starting from the vector-axial Lagrangian of Eq.~\eqref{eq:toylagrangian}, let us perform now an $SU(N)$ vector fermion reparametrisation,
\begin{equation}
	\psi \to e^{i\theta(x)} \psi, \quad \bar\psi \to \bar \psi\, e^{-i\theta(x)} \ .
\end{equation}
Under this fermion reparametrisation, the Lagrangian given by Eq.~\eqref{eq:toylagrangian} becomes,
\begin{equation}
	\mathcal L \to \bar \psi \big[ i\slashed D - m - \big(\slashed D \theta\big) \big] \psi 	\ ,
\end{equation}
with $D_\mu \equiv \big( \partial_\mu + iV_\mu + iA_\mu\gamma_5 \big)$, and again the quantity inside the parenthesis, $\big(\slashed D\theta\big) \equiv \gamma^\mu(\partial_\mu \theta+i[V_\mu,\theta]+i[A_\mu,\theta]\gamma_5)$, indicates that the covariant derivative locally acts on $\theta(x)$, with $\theta$ charged under the gauge group. The Jacobian produced by this transformation is therefore given by the following expression,
\begin{equation}
	J[\theta] = \dfrac{\det \big(i \slashed D -m\big)}{\det\big(i\slashed D -m - (\sD \theta)\big)} \ .
	\label{Jacobian: vector-global-rotation}
\end{equation}
As in the axial rotation, the functional determinants of Eq.~\eqref{Jacobian: vector-global-rotation} are ill-defined and need to be regularised. In dimensional regularisation, this is the $\gamma_5$ located in the covariant derivative which entirely carries the ambiguity now.

Before presenting the computation of the Jacobian of Eq.~\eqref{Jacobian: vector-global-rotation} and its bosonised form, one should notice that starting from Eq.~\eqref{covabos}, no additional computation is needed, if one is only interested in the result. Indeed, from
\begin{equation}
	\partial_\mu J^\mu_5 = \partial_\mu J^\mu_R - \partial_\mu J^\mu_L = \dfrac{-i}{32\pi^2}\epsilon^{\mu\nu\rho\sigma}\tr\,\theta(F^L_{\mu\nu}F^L_{\rho\sigma}+F^R_{\mu\nu}F^R_{\rho\sigma}) \ ,
\end{equation}
one can identify,
\begin{equation}
	\partial_\mu J_R^\mu = \dfrac{-i}{32\pi^2}\epsilon^{\mu\nu\rho\sigma}\tr\,\theta(F^L_{\mu\nu}F^L_{\rho\sigma}), \quad \partial_\mu J^\mu_L = \dfrac{i}{32\pi^2}\epsilon^{\mu\nu\rho\sigma}\tr\,\theta(F^R_{\mu\nu}F^R_{\rho\sigma}) \ ,
\end{equation}
with $\theta\in SU(N)$.
Hence,
\begin{align}
\partial_\mu J^\mu_V &= \partial_\mu J^\mu_R + \partial_\mu J^\mu_L 
= \dfrac{-i}{32\pi^2}\epsilon^{\mu\nu\rho\sigma}\tr\,\theta \big(F^R_{\mu\nu}F^R_{\rho\sigma} - F^L_{\mu\nu}F^L_{\rho\sigma}) \nonumber \\
&= -\frac{i}{16\pi^2}\epsilon^{\mu\nu\rho\sigma}\tr\,\theta(F^V_{\mu\nu}F^A_{\rho\sigma}+F^A_{\mu\nu}F^V_{\rho\sigma}) \ .
\end{align}
We are however more interested in presenting an explicit and transparent evaluation of this version of the covariant anomaly.

\paragraph{Fermionic expansion with free parameters}

We proceed in a similar fashion as for the covariant anomaly in the axial current, except there is no mass term to compute according to Eq.\eqref{Jacobian: vector-global-rotation}.

The derivative term is ambiguous because of the presence of $\gamma_5$ in the covariant derivative, and of the divergent integrals. As explained in section \ref{subsec:axial-global-anomalous}, it can thus be written as,
\begin{equation}
\mathcal{A} = 
\mathcal{A}_{\slashed\partial}=\frac{-i}{16\pi^2}\epsilon^{\mu\nu\rho\sigma}\tr\,\theta^a T^a\left(\sum_i a_i X_{i,\,\mu\nu\rho\sigma} \right)\, ,
\end{equation}
where $X_i$ are all the possible operators of the form $\mathcal{O}_1\mathcal{O}_2\mathcal{O}_3\mathcal{O}_4$ with $\mathcal{O}_{1\leq i\leq4}\in\{V,A,\pa\}$ that can be formed. Contrary to the case of the anomaly in the axial current, the operators $X_i$ that appear now have an odd number of $A$ (because there must be an odd number of $\gamma_5$). Note that the operators with a partial derivative to the right vanish, and those with consecutive partial derivatives vanish due to the contraction with the $\epsilon$ tensor. This leaves us with again 22 possible operators, with 22 free parameters $a_i$.

The result should be covariant under the gauge transformation Eq.~\eqref{gaugevarVA}. Once again, we only need to enforce the gauge covariance with respect to $\epsilon_A$ to fix the free parameters. The covariance of the result requires that its gauge variation has the form Eq.~\eqref{GaugeCov}, we therefore enforce on the free parameters that the derivatives of $\epsilon_A$ and the operators that have $\epsilon_A$ neither at the beginning nor at the end of the operator vanish. This fixes again 21 free parameters out of 22, and leaves us with a result of the form,
\begin{equation}
\mathcal{A}_{\slashed\partial} = 
\alpha\, \epsilon^{\mu\nu\rho\sigma}\tr\,\theta\left( F^V_{\mu\nu}F^A_{\rho\sigma}+F^A_{\mu\nu}F^V_{\rho\sigma} \right)\, ,
\label{cov-anomaly: vector-unfixed}
\end{equation}
where $\alpha$ is the remaining free parameter.

For the anomaly in the axial current, we compared our result to the ABJ anomaly by setting the gauge field $A=0$ to fix the normalisation. Unfortunately, this is not possible here because setting $A=0$ makes the whole term vanish. 

In the case of the Abelian anomaly, we happen to have the same issue, where the result is gauge covariant but there is a normalisation freedom that remains. In Ref.~\cite{Quevillon:2021sfz}, they deal with the free parameters for the Abelian anomaly to fix the normalisation factor by enforcing the conservation of the axial current (up to the mass term)\footnote{For the reader interested in anomaly from the global axial(vector) transformation, see Ref.~\cite{Quevillon:2021sfz} for the detail of computations and also the applications in axion phenomenology.}. In Eq.~\eqref{cov-anomaly: vector-unfixed}, we set the gauge fields and $\theta$ as Abelian\footnote{Eq.~\eqref{cov-anomaly: vector-unfixed} can be separated in an Abelian part, and a non-Abelian part (formed uniquely of commutators of $\theta$, $V$ and $A$). Since the free parameter is common to both these parts, we can set the non-Abelian part to zero to fix the free parameter as in the Abelian case.} and apply the technique from Ref.~\cite{Quevillon:2021sfz}.

We consider Eq.~\eqref{cov-anomaly: vector-unfixed} with Abelian gauge fields and Abelian $\theta$, which is gauge covariant independently of the remaining free parameter. To break down this gap, we re-organise Eq.~\eqref{cov-anomaly: vector-unfixed} in terms of Generalised Chern-Simons (GCS) forms using integration by parts, then we introduce an auxiliary background field $\xi_{\mu}$ associated to the deformation of $(\partial_{\mu} \theta)$\,\footnote{The auxiliary vector field $\xi_{\mu}$ will be set to zero at the end of the computation.} as follows,
\begin{align}
    \mathcal{A}_{\slashed\partial} &=
    \beta\,\epsilon^{\mu\nu\rho\sigma}\tr\big[\xi_{\mu} - (\partial_{\mu}\theta) \big](iA_{\mu}) F^V_{\rho\sigma}
    \, ,
    \label{cov-anomaly: vector-GCS}
\end{align}
where $\beta=4\alpha$. At this stage, Eq.~\eqref{cov-anomaly: vector-GCS} is no longer gauge invariant under the axial gauge transformation. The conservation of the axial current (up to the mass term) can be enforced non-trivially if the axial gauge field obtains its mass after spontaneous symmetry breaking. By introducing the Goldstone boson $\pi_A$ associated to the longitudinal component of the axial gauge field $A_{\mu}$, we obtain,
    \begin{align}
        \tilde{\mathcal{A}}_{\slashed\partial} 
        &= \beta\,\epsilon^{\mu\nu\rho\sigma}\tr\big[\xi_{\mu} - (\partial_{\mu}\theta) \big](iA_{\mu}) F^V_{\rho\sigma}
        -\dfrac{i}{8\pi^2}\,\epsilon^{\mu\nu\rho\sigma}\,\tr\bigg[\dfrac{\pi_A}{v} (\partial_{\mu}\xi_{\nu})F^V_{\rho\sigma} \bigg]
        \, .
    \end{align}
Requiring the quantity $\tilde{\mathcal{A}}_{\slashed\partial}$ to be gauge invariant, implies that $\beta=-i/(4\pi^2)$, or equivalently $\alpha = -i/(16\pi^2)$. Additional details about the GCS terms and the Goldstone terms are provided in the appendix~\ref{appfreeparam}. Eventually, going back to non-Abelian gauge fields and $\theta$, we obtain the non-Abelian covariant anomaly in the vector current,
\begin{align}
    \mathcal{A}_{\slashed\partial} &
    =\dfrac{-i}{16\pi^2}\epsilon^{\mu\nu\rho\sigma}\, \tr \, \theta \big( F_{\mu\nu}^V F^A_{\rho\sigma}+F_{\mu\nu}^A F^V_{\rho\sigma} \big)
    \, .
    \label{cov-anomaly: vector-fixed}
\end{align}

\paragraph{Bosonisation method.} \label{covbosformvec}

We bosonise the Jacobian from Eq.~\eqref{Jacobian: vector-global-rotation}. Following the method detailed in Eq.~\eqref{jacboson}
\begin{align}
\begin{split}
J[\theta]^2=\frac{\det (-(i\slashed D)^\dagger i\sD+m^2)}{\det (-(i\sD)^\dagger i\sD+m^2+m(i\sD-(i\sD)^\dagger)+f(\theta))} \ ,
\label{jacbosonVecTransfo}
\end{split}
\end{align}
and we have defined,
\begin{align}
\begin{split}
f(\theta)=i[\theta,D^2]-\half[\sigma.F^V,\theta]-\half[\sigma.F^A\gamma_5,\theta]\ .
\end{split}
\end{align}
However, this regularisation yields the covariant anomaly in the axial current as seens in the above, it thus comes with no surprise that the vector current is conserved. That is to say, if we supplement the theory with a global axial and a global vector symmetries, then this regularisation puts all the anomaly in the global axial symmetry and conserves the global vector symmetry. Therefore, as expected, the Jacobian in Eq.~\eqref{jacbosonVecTransfo} is equal to one and then is unable, from the start, to deal with an anomalous vector transformation.

\subsection{Consistent anomaly}

For the covariant anomaly, we have showed that we can bosonise the Jacobian in a gauge covariant way. However,  the anomalous operator $\mathcal{A}$ may not be gauge invariant and in that case, one should rather make sure that the determinants in Eq.~(\ref{JacGeneral}) are regularised in a non-gauge invariant way. Now, what we may ask is that the anomaly satisfies the algebra of the gauge group i.e the anomaly can be required to satisfy a consistency relation (also called an integrability condition or Wess-Zumino condition~\cite{Wess:1971yu}). In that case, the anomaly is more accurately called the consistent anomaly.

As a  remark, it has been shown that the Wess-Zumino condition corresponds to the Bose symmetry with respect to the vertices of the one-loop Feynman diagrams. If the covariant anomaly collects the effects of the anomaly to only one of the vertices, this does not satisfy the Bose symmetry and thus the so-called integrability condition. Notice that the leading terms of consistent anomaly and covariant anomaly (e.g the anomaly corresponding to the $AAA$ triangle diagrams) are related by the Bose symmetry factor. In 4 dimensions, the symmetry factor is $1/3$. For arbitrary $2n$ dimensions, the symmetry factor is $1/(n+1)$. The reason for these symmetry factors lies in the distribution of anomaly in all vertices when evaluating the consistent anomaly\footnote{We remind the reader that the Bose symmetry will play an essential role in the functional bosonisation formalism; for further discussions, see Ref.~\cite{Banerjee:1995np}.}.

\subsubsection{Fermionic expansion with free parameters}

As it is done in the previous sections, it is possible to compute the anomaly without bosonising the Jacobian, although it is ambiguous. This ambiguity transpires in certain traces that bear a $\gamma_5$ in $d$ dimensions. Keeping track of the ambiguity requires the introduction of free parameters that need to be fixed under physical constraints. For the covariant anomaly, those physical constraints arise from the expected gauge covariance of the result. However, for the consistent anomaly, it is not gauge covariance or invariance that needs to be enforced, but rather Wess-Zumino consistency conditions.
We will outline the method in the following.

The covariant derivative is $i\sD = i\slashed \partial - \slashed V - \slashed A \gamma_5$. Under an axial reparametrisation of the fermions, the path integral yields the following Jacobian,
\begin{equation}
	J[\theta] 
	= \dfrac{\det(i\slashed D -m)}{\det\big[ e^{i\theta\gamma_5}\big( i\slashed D -m \big)e^{i\theta\gamma_5} \big]}
	=\frac{\det(i\slashed D -m)}{\det(i\slashed D -m -(\slashed D\theta)\gamma_5-2im\theta \gamma_5)}\, ,
\end{equation}
where $\theta=\theta^a T^a$ is charged under the gauge group $SU(N)$.

The anomalous operator can be expressed as the following expansion,
\begin{equation}
\mathcal A= - \int \frac{\dd^d q}{(2\pi)^d}\tr\,(-2im\theta \gamma_5-(\slashed D\theta)\gamma_5) \sum_{n\geq0}\left[\left(\frac{-1}{\slashed q+m} \right)(-i\slashed D)\right]^n\left(\frac{-1}{\slashed q+m} \right)\, .
\end{equation}
As we will see, the mass term $2im\theta\gamma_5$ gives rise to the anomaly, while the divergent term $(\slashed D\theta)\gamma_5$ does not contribute to the result at order $m^0$. However the derivative term will contribute at higher order to cancel the contributions from the mass term, so that the whole result is proportional to $m^0$.\\

The computation is the same as the one of the consistent anomaly in section \ref{FermExpFreeParam}. We thus have,
\begin{align}
\begin{split}
\mathcal{A}^{m\gamma_5}=\frac{-i}{16\pi^2}\epsilon^{\mu\nu\rho\sigma}\tr\, \theta^aT^a\bigg(& F^V_{\mu\nu}F^V_{\rho\sigma}+\frac{1}{3}F^A_{\mu\nu}F^A_{\rho\sigma}\\
&-\frac{8}{3}\left( iA_\mu iA_\nu F^V_{\rho\sigma} +iA_\mu F^V_{\nu\rho} iA_\sigma +F^V_{\mu\nu}iA_\rho iA_\sigma \right) +\frac{32}{3} iA_\mu iA_\nu iA_\rho iA_\sigma \bigg)\\
&\hspace{-3.7cm}=\mathcal{A^{\mathrm{Bardeen}}}\, ,
\label{BardeenAn}
\end{split}
\end{align}
and,
\begin{equation}
\mathcal{A}_{\slashed\pa\gamma_5}=\frac{-1}{16\pi^2}\epsilon^{\mu\nu\rho\sigma}\tr\,\theta^a T^a\left(\sum_i a_i {X_i}_{\mu\nu\rho\sigma} \right)\, ,
\end{equation}
where $X_i$ are all the possible operators of the form $\mathcal{O}_1\mathcal{O}_2\mathcal{O}_3\mathcal{O}_4$ with $\mathcal{O}_{1\leq i\leq4}\in\{V,A,\pa\}$ that can be formed, provided it has an even number of $A$ fields (the number of $\gamma_5$ must be odd). Note that the operators with a partial derivative to the right vanish, and those with consecutive partial derivatives vanish due to the contraction with the $\epsilon$ tensor. This leaves us with 22 possible operators, with 22 free parameters $a_i$.

Let's take all the operators from the derivative term when the axial field $A$ goes to zero. There remains only the operators that do not depend on $A$ (they only bear $V$ and $\pa$) and each have a free parameter. Setting $A$ to zero amounts to considering an axial reparametrisation of the fermion in a vector gauge theory, and if we want for example to conserve the vector current we know the result should then be vector gauge invariant. Therefore, the free parameters are fixed under this requirement, and the terms that do not depend on $A$ combine together to form,
\begin{equation}
\alpha\frac{-i}{16\pi^2}\epsilon^{\mu\nu\rho\sigma}\tr\,\theta \left.F^V_{\mu\nu}\right|_{A=0}\left.F^V_{\rho\sigma}\right|_{A=0}\, ,
\end{equation}
with $\left.F^V_{\mu\nu}\right|_{A=0}=i\left((\pa_\mu V_\nu)-(\pa_\nu V_\mu)+i[V_\mu,V_\nu]\right)$, and $\alpha$ is a remaining free parameter that cannot be fixed by the sole requirement of gauge invariance (while $A=0$).

We rewrite this term as,
\begin{align}
\begin{split}
&\alpha\frac{-i}{16\pi^2}\epsilon^{\mu\nu\rho\sigma}\tr\,\theta \left.F^V_{\mu\nu}\right|_{A=0}\left.F^V_{\rho\sigma}\right|_{A=0}\\
&=\alpha\frac{-i}{16\pi^2}\epsilon^{\mu\nu\rho\sigma}\tr\,\theta\bigg[ F^V_{\mu\nu}F^V_{\rho\sigma}\\
&\quad\quad\quad\quad- \left.F^V_{\mu\nu}\right|_{A=0}i^2[A_\rho,A_\sigma]-i^2[A_\mu,A_\nu]\left.F^V_{\rho\sigma}\right|_{A=0}-i^2[A_\mu,A_\nu]i^2[A_\rho,A_\sigma]\bigg]\, ,
\label{FVA=0toFV}
\end{split}
\end{align}
where we have made the Bardeen curvature of Eq.~\eqref{eq:FV} appear.\\

Now consider the remaining terms with free parameters, that is to say, the terms that vanished when we set $A$ to zero. Among those operators, we can identify the same operators as those in the last line of equation Eq.~\eqref{FVA=0toFV}, with different free parameters $\beta_i$. Therefore, they will combine together, and only change the free parameters $\beta_i$ to new free parameters $\beta_i'$.\\

We will now enforce the Wess-Zumino consistency conditions. No calculation is needed, we will only use the well known fact that the Wess-Zumino consistency conditions fix the coefficients of all the operators with respect to the coefficient of the term $F^V_{\mu\nu}F^V_{\rho\sigma}$ (as explained in Ref.~\cite{Wess:1971yu}). Therefore, among all the remaining operators, all the free parameters will be fixed with respect to one, $\alpha$ in Eq.~\eqref{FVA=0toFV}, such that the whole operator respects the integrability conditions. This unequivocally leaves us with,
\begin{align}
\begin{split}
\mathcal{A}^{\slashed\pa \gamma_5}=\alpha\frac{-i}{16\pi^2}\epsilon^{\mu\nu\rho\sigma}\tr\, \theta^aT^a\bigg(& F^V_{\mu\nu}F^V_{\rho\sigma}+\frac{1}{3}F^A_{\mu\nu}F^A_{\rho\sigma}\\
&-\frac{8}{3}\left( iA_\mu iA_\nu F^V_{\rho\sigma} +iA_\mu F^V_{\nu\rho} iA_\sigma +F^V_{\mu\nu}iA_\rho iA_\sigma \right) +\frac{32}{3} iA_\mu iA_\nu iA_\rho iA_\sigma \bigg)\\
&\hspace{-4.0cm} = \alpha\mathcal{A}^{\mathrm{Bardeen}}\, ,
\label{alphaBardeenAn}
\end{split}
\end{align}
where we still have one free parameter left, $\alpha$.

\paragraph{Final result.}

Now let's put together the contributions from the mass term and the derivative term, i.e Eqs.~\eqref{BardeenAn} and \eqref{alphaBardeenAn} , we obtain,
\begin{equation}
\mathcal{A}=(1+\alpha)\mathcal{A^{\mathrm{Bardeen}}}\, .
\end{equation}
Basically, the Wess-Zumino consistency conditions allow us to fix the coefficients of all the operators with respect to the coefficient of the term $F^V_{\mu\nu}F^V_{\rho\sigma}$, this is why we still have a remaining freedom at the end. The coefficient of $F^V_{\mu\nu}F^V_{\rho\sigma}$ can be fixed by comparing the result with the anomaly in a vector gauge theory as suggested in \cite{Wess:1971yu}. That is to say, in our result, we set again $A$ to zero, therefore we can identify our result with the ABJ anomaly (with $\theta\in SU(N)$) of Eq.~\eqref{ABJanomaly}, which immediately sets $\alpha$ to zero, leaving the expected result.

Note that there is no need to introduce counter terms in our computation to obtain the minimal Bardeen anomaly, because the vector current conservation has been enforced to fix the free parameters.

As a remark, notice that by comparing our result with the ABJ anomaly to fix the last free parameter, we restrain ourselves to the consistent anomaly with the vector current being conserved. If we want for example to conserve the axial current, we need to compare with the anomaly in a vector gauge theory where the anomaly is in the vector current. Besides, the Wess-Zumino consistency conditions have to be adapted. Indeed, they correspond to enforcing the Lie algebra of the gauge group and the Ward identities as well. Changing the current that remains conserved at the quantum level amounts to changing the Ward identities, hence changing the Wess-Zumino consistency conditions.

The procedure presented in this section can thus also be applied while enforcing the conservation of the axial current. It can even be used to obtain a generic expression where the consistent anomaly is distributed between the vector and the axial currents. 

As far as we know, this is the only method that allows to tune which current bears the anomaly from a path integral approach.

\paragraph{Calculation in BMHV's scheme.}

Alternatively, it is possible to obtain the Bardeen anomaly without relying on free parameters. It is known that Pauli-Villars regularisation satisfies the Wess-Zumino consistency conditions, as well as enforcing conservation of the vector current \cite{Fujikawa:2004cx}. Besides, as showed in Refs.~\cite{Horejsi:1988ei,Novotny:1994yx,Bertlmann:1996xk}, BMVH scheme in dimensional regularisation is equivalent to a "continuous superposition" of Pauli-Villars regularisations, and thus respects the Wess-Zumino consistency conditions, and vector current conservation as well. We can therefore avoid the introduction of free parameters and significantly simplify the calculation by making use of BMHV scheme to obtain Bardeen's consistent anomaly in the axial current. Although it strips us of the freedom to chose which current should bear the anomaly, as opposed to the free parameters approach discussed above.

\subsubsection{Bosonisation method}
\label{consano}

The bosonisation presented in section \ref{covbosform} defines a finite and non-ambiguous Jacobian. But it only allows us to get a gauge covariant result. The same procedure thus cannot be used to compute the consistent anomaly.

Nonetheless, we can try to bosonise with the operator $\sD^2$. It has the same spectrum as $i\sD$ (aside squaring it). We circumvent the problem of the non-hermitianity using the analytic continuation $A_\mu\to iA_\mu$ \cite{Bertlmann:1996xk,Fujikawa:2004cx}.

We showed that bosonising using $(i\sD)^\dagger i\sD$ as a regulator enforces the gauge covariance of the result, leaving us with only the possiblity to get the covariant anomaly. However, there is no reason to think that bosonising with the analytic continuation $A_\mu \to iA_\mu$ and $\sD^2$ would enforce all the conditions to get the consistent anomaly, namely the  Wess-Zumino (integrability) consistency conditions.\\

Let's now see in the computation why the $\sD^2$ bosonisation along with the analytic continuation is still ambiguous.

After the analytic continuation, we have $i\sD = i\slashed \partial - \slashed V - i\slashed A \gamma_5$. The Jacobian of the axial field reparametrisation is the following,
\begin{equation}
	J[\theta] 
	=\frac{\det (i\sD-m)}{\det\left( e^{i\theta\gamma_5}(i\sD-m)e^{i\theta\gamma_5}\right)}=\frac{\det(i\slashed D -m)}{\det(i\slashed D -m -2im\theta\gamma_5-(\sD\theta) \gamma_5)} \ ,
\end{equation}
where $\theta=\theta^a T^a$ is charged under the $SU(N)$ gauge group.

Now we perform the bosonisation,
\begin{align}
\begin{split}
J^2[\theta] &=\frac{\det (i\sD-m)}{\det\left( e^{i\theta\gamma_5}(i\sD-m)e^{i\theta\gamma_5}\right)}\frac{\det (i\sD-m)}{\det\left( e^{i\theta\gamma_5}(i\sD-m)e^{i\theta\gamma_5}\right)}\\
&=\frac{\det (i\sD-m)}{\det\left( e^{i\theta\gamma_5}(i\sD-m)e^{i\theta\gamma_5}\right)}\frac{\det (-i\sD-m)}{\det\left( e^{i\theta\gamma_5}(-i\sD-m)e^{i\theta\gamma_5}\right)}\\
&=\frac{\det(\sD^2+m^2)}{\det(\sD^2+m^2+[2im\theta\gamma_5,i\sD]+\{i\sD,(\sD\theta)\gamma_5\}+4im^2\theta\gamma_5)} \ ,
\label{jacbosonCons}
\end{split}
\end{align}
where we have use the fact that the determinant is invariant under the change $\gamma^\mu\to -\gamma^\mu$ (see footnote \ref{footnoteSignFlip}). We expand it following the prescription described in section~\ref{sec:Main},
\begin{equation}
\log J[\theta]^2=\int\dd^4x \frac{\dd^d q}{(2\pi)^d}e^{-iq\cdot x}\tr\, \left([2im\theta\gamma_5,i\sD-\slashed q]+\{i\sD-\slashed q,(\sD\theta)\gamma_5\}+4im^2\theta\gamma_5\right)\frac{1}{\sD^2+m^2} e^{iqx} \ .
\end{equation}
We can straightforwardly see that the term $[2im\theta\gamma_5,i\sD-\slashed q]/(\sD^2+m^2)$ has an odd number of Dirac matrices, therefore it vanishes under the trace. 
For simplicity we extract the $\gamma_5$ in $D_\mu$ using the notations,
\begin{equation}
iD_\mu=iD^V_\mu-iA_\mu\gamma_5\,\text{ where }\,iD^V_\mu=i\pa_\mu-V_\mu \ .
\end{equation}
We have,
\begin{align}
\begin{split}
e^{-iq\cdot x}\sD^2e^{iq\cdot x}&=(\sD+i\slashed q)(\sD+i\slashed q)\\
&=\sD^2-q^2+2iq\cdot D^V-[\gamma^\mu,\gamma^\nu]\gamma_5iq_\mu A_\nu \ .
\end{split}
\end{align}

\noindent
Finally, we can expand the Jacobian as usual,
\begin{equation}
2\mathcal{A}=-\int\frac{\dd^4q}{(2\pi)^4}\tr\,\left(\{i\sD-\slashed q,(\sD\theta)\gamma_5\}+4im^2\theta\gamma_5\right)\sum_{n\geq0}\left[\Delta(\sD^2+2iq\cdot D^V-[\gamma^\mu,\gamma^\nu]\gamma_5iq_\mu A_\nu)\right]^n\Delta  \ ,   
\end{equation}
where $\Delta=1/(q^2-m^2)$.\\
The key point is that because of the term $-[\gamma^\mu,\gamma^\nu]\gamma_5iq_\mu A_\nu$, there are terms with several momenta $q$ contracted with a Dirac matrix as $\gamma^\mu q_\mu$. Combined with divergent integrals, they lead to traces traces such as,
\begin{equation}
    \tr\,(\gamma^\mu\gamma^\nu\gamma^\rho\gamma^\sigma\gamma^a\gamma_5\gamma_a)\, ,
\end{equation}
which are ambiguous in $d$ dimensions. When bosonising using $(i\sD)^\dagger i\sD$, the term $-[\gamma^\mu,\gamma^\nu]\gamma_5iq_\mu A_\nu$ does not appear (see Eqs.~\eqref{DdagDFourier1} and \eqref{DdagDFourier2} in the appendix~\ref{appbosonized}).

Therefore, the bosonisation method does not offer any appealing simplification regarding the calculation of the consistent anomaly. One could of course, proceed with the computation with free parameters or get rid of those ambiguities using the BMHV scheme which satisfies the Wess-Zumino conditions but also enforces vector current conservation. We refrain from doing so as ultimately, this does not offer any insights compared to the fermionic expansion already discussed.

\section{Axial-gravitational anomaly}

In this section we aim at deriving the axial-gravitational anomaly, which stems from the gravitational contribution to the Jacobian associated with the axial reparametrisation as defined in Eq.~\eqref{FermionRep: axial}. 

In curved space-time, the covariant derivative does not only bear the gauge fields. Diffeomorphism invariance requires the presence of the Christoffel connection, and Lorentz invariance requires the presence of the spin-connection for fermions.
For simplicity, we consider a theory without gauge sector (in any case we know that we do not expect cross terms between the gravity sector and the gauge sector). We denote by $D_\mu$ the general covariant derivative, it includes both the spin-connection when applied to non-trivial element of the Dirac space, and the Christoffel symbols when applied to a Lorentz tensor. We follow \cite{Parker:2009uva} for conventions for the spin-connection. We have,
\begin{equation}
D_\mu\Psi=(\partial_\mu+\omega_\mu)\Psi \, ,
\end{equation}
with the the spin-connection defined as $\omega_\mu=\frac{1}{8}[\gamma^a,\gamma_b] e^\nu_a(D_\mu e^b_\nu)$, with $e^a_\mu$ the tangent frame vielbein such that $g_{\mu\nu}=e^a_\mu e^b_\nu g_{ab}$ (latin indices referring to the tangent frame).

Considering a spinless Lorentz vector $v$ we have,
\begin{equation}
D_\mu v^\nu =\partial_\mu v^\nu+\Gamma^\nu_{\mu\rho}v^\rho
\end{equation}
\begin{equation}
D_\mu v_\nu =\partial_\mu v_\nu-\Gamma^\rho_{\mu\nu}v_\rho \, .
\end{equation}
The following expressions will be useful later,
\begin{align}
&F_{\mu\nu}\Psi=[D_\mu,D_\nu]\Psi=\frac{1}{4}\gamma^\rho\gamma^\sigma R_{\mu\nu\rho\sigma}\Psi 
\label{eq:Fgrav}\\
&\sD^2=D^2-\frac{i}{2}\sigma^{\mu\nu}F_{\mu\nu} \text{ where } \sigma^{\mu\nu}=\frac{i}{2}[\gamma^\mu,\gamma^\nu] \, ,
\label{eq:sP2}
\end{align}
with
\begin{equation}
\frac{i}{2}\sigma.F\,\Psi=\frac{1}{4}R\, \mathds{1}_{\mathrm{Dirac}}\Psi \, ,
\label{eq:sigmaFgrav}
\end{equation}
where $\mathds{1}_{\mathrm{Dirac}}$ is the identity in Dirac space.

Finally, the covariant derivative commutes with the Dirac matrices,
\begin{align}
(D_\mu\gamma^\nu)&=(\pa_\mu\gamma^\nu)+\Gamma^\nu_{\mu\rho}\gamma^\rho+[\omega_\mu,\gamma^\nu]=0\\
(D_\mu \gamma_5)&=[\omega_\mu,\gamma_5]=0 \, .
\end{align}

\subsection{Covariant Derivative Expansion in curved space-time}

The CDE in curved space-time requires extra care that significantly complexifies the expansion.
The main point is that the commutativity between the covariant derivatives $D$ and the propagators $\Delta=1/(q^2-m^2)$ that appear in our expansion is lost. Indeed, $q^2=g^{\mu\nu}(x)q_\mu q_\nu$ is space-time dependent. Therefore $[D_\mu,\Delta]=-(\partial_\mu q^2)\Delta^2$.
The CDE can still be performed in an extended framework that includes the curvature of space-time, as it has been done for example in Refs.\cite{Binetruy:1988nx,Alonso:2019mok}.
However we propose here a different way of conducting the expansion, that we believe to be simpler in the formalism. Defining the expansion in curved space-time is not trivial, and this is out of the scope of this paper (see Ref.~\cite{LarueQuevillon}), so we just give the outline of the method without delving too much into the details.

First of all, there is no trivial definition of the Fourier transform in curved space-time. However, following Refs.~\cite{Bunch:1979uk,Alonso:2019mok} we can define the Fourier transform using Riemann Normal Coordinates (RNC). We take the momentum $q_\mu$ to be the covariant variable conjugate to the contravariant variable $x^\mu$, so that $\dd^D q\dd^Dx$ is diffeomorphism-invariant. We then have,
\begin{equation}
(\partial_\mu q_\nu)=\frac{\partial q_\nu}{\partial x^\mu}=0\, ,
\end{equation}
but,
\begin{equation}
(\partial_\mu q^\nu)=(\partial_\mu g^{\nu\rho}q_\rho)=(\partial_\mu g^{\nu\rho})q_\rho\neq0 \, .
\end{equation}
We thus have the standard Fourier transform of the covariant derivative,
\begin{equation}
e^{-iq\cdot x}D_\mu e^{iq\cdot x}=D_\mu+(\partial_\mu iq_\nu x^\nu)=D_\mu+iq_\nu(\partial_\mu x^\nu)=D_\mu+iq_\mu \, .
\label{FourierGrav}
\end{equation}

\subsection{Computation of the gravitational anomaly}

According to the previous results, we know that at order $m^0$ the derivative coupling does not contribute in the bosonised form (and we can show it ), for simplicity we drop it. The anomaly is thus fully encompassed (at order $m^0$) by the following Jacobian~\footnote{We have decided to work within the bosonised form of the Jacobian, but one could have equivalently chosen to carry the computation with the original Jacobian.},
\begin{equation}
J[\theta]^2=\frac{\det\left(\g^2(\sD^2+m^2)\right)}{\det\left(\g^2(\sD^2+m^2+4im^2\theta\gamma_5 )\right)} \, .
\end{equation}
Since we discarded the derivative term, this Jacobian is trivially finite, thus well-defined.

Using Eq.~\eqref{FourierGrav}, we have,
\begin{align}
\begin{split}
e^{-iq\cdot x}\sD^2 e^{iq\cdot x}&=g^{\mu\nu}(D_\mu+iq_\mu)(D_\nu+iq_\nu)-e^{-iq\cdot x}\frac{i}{2}\sigma.F e^{iq\cdot x}\\
&=D^2-q^2+2iq\cdot D+ig^{\mu\nu}(D_\mu q_\nu)-\frac{i}{2}\sigma.F\\
&=D^2-q^2+2iq\cdot D-i\Gamma^\rho_{\mu\nu}q_\rho-\frac{i}{2}\sigma.F  \, .
\end{split}
\end{align}
Hence we can expand the Jacobian as,
\begin{equation}
\log J[\theta]^2=\int \dd^4 x\frac{\dd^d q}{(2\pi)^d}
\tr\left( 4im^2\theta\gamma_5\sum_{n\geq0}\left[\Delta(D^2-\frac{i}{2}\sigma.F+2iq\cdot D-i\Gamma^\rho_{\mu\nu}q_\rho)\right]^n\Delta\right) \, .
\label{JacGrav2}
\end{equation}
Since all the Lorentz indices are contracted, the field strength that appears in Eq.~\eqref{JacGrav2} is in the fermion representation\footnote{Indeed, recall that the trace in internal space (\textit{ie} Dirac space and gauge space) is defined as $\tr A=\sum_{n}\Psi_n^\dagger A \Psi_n$, where $\{\Psi_n\}$ is a basis of internal space (constant vectors: $(D_\mu\Psi_n)=\omega_\mu\Psi_n+V_\mu\Psi_n$). Therefore, for any operator $\mathcal{O}$ that is a matrix in internal space without free Lorentz indices (and it can bear open covariant derivatives) we have: $\tr D_\mu\mathcal{O}=\sum_n\Psi_n^\dagger D_\mu\mathcal{O}\Psi_n$. Since $\mathcal{O}$ is a matrix in internal space, then $\mathcal{O}\Psi_n$ is a vector in internal space, and all the derivatives in $\mathcal{O}$ are localised because they act on $\Psi_n$. Therefore, in $\sum_n\Psi_n^\dagger D_\mu\mathcal{O}\Psi_n$, $D_\mu$ acts on a vector in internal space, hence can be written in the fermion representation.}, hence we can use Eq.~\eqref{eq:sigmaFgrav} to simplify,
\begin{equation}
\log J[\theta]^2=\int \dd^4 x \frac{\dd^d q}{(2\pi)^d}
\tr\left( 4im^2\theta\gamma_5\sum_{n\geq0}\left[\Delta(D^2-\frac{R}{4}+2iq\cdot D-i\Gamma^\rho_{\mu\nu}q_\rho)\right]^n\Delta\right) \, .
\label{JacGrav}
\end{equation}
We remark that the space-time measure $\g$ does not come into play in the expansion. This is because it appears both in the numerator and the denominator.\\

Now, we are interested in the terms that are proportional to $m^0$. In each of these terms, the propagators $\Delta=1/(q^2-m^2)$ have to be commuted to the left in order to perform the integration over momentum. Therefore, each of these terms will yield several terms where the open covariant derivatives will be localised or not on a propagator.

For example, consider the following term of order $m^0$,
\begin{align}
&\int \frac{\dd^d q}{(2\pi)^d} \tr\left[4im^2\theta\gamma_5\Delta\left(-\frac{R}{4}\right)\Delta D^2\Delta\right]
\nonumber \\
&=\int \frac{\dd^d q}{(2\pi)^d}\Delta^2 \tr\left[4im^2\theta\gamma_5\left(-\frac{R}{4}\right) (\Delta D^2+(D^2\Delta)+2(D^\mu\Delta)D_\mu)\right] \, .
\label{GravExample}
\end{align}
The only terms that can contribute in the end are gauge and diffeomorphism invariant. That is to say the remaining open covariant derivatives that are not localised on a propagator have to combine together to form field strengths. For example in Eq.~\eqref{GravExample}, the term involving $(D_\mu\Delta)D_\nu$ has a single open derivative, it is impossible to form an invariant term with it, thus it cannot contribute to the final result (besides it vanishes in Riemann Normal Coordinates).

Secondly, notice that whenever a covariant derivative is localised on a propagator, it bears no spin-connection since $\Delta$ is a scalar in Dirac space: $(D_\mu\Delta)=(\pa_\mu\Delta)\mathds{1}_{\mathrm{Dirac}}$.

Keeping those last two points in mind, one can easily isolate the few terms that will contribute to the gravitational anomaly, making use of,
\begin{align}
\tr\,\gamma_5=\tr\,\gamma_5\gamma^\mu\gamma^\nu &=0 
\, , \quad 
\tr\,\gamma_5\gamma^\mu\gamma^\nu\gamma^\rho\gamma^\sigma \neq0 \, .
\end{align}
In Eq.~\eqref{JacGrav}, the only terms that bear Dirac matrices are the covariant derivatives via the spin-connection. The remaining open covariant derivatives will combine in the end to form field strengths that have two Dirac matrices (see Eq.~\eqref{eq:Fgrav}), therefore the only way to have enough Dirac matrices so that the trace does not vanish is by having 4 open covariant derivatives, thus two field strengths.

In the end, the only terms that can contribute are the following,
\begin{itemize}
    \item At $n=2$: $\int \frac{\dd^d q}{(2\pi)^d}\Delta^3 \tr\left[4im^2\theta\gamma_5 D^2 D^2\right]$
    \item At $n=3$: $\int \frac{\dd^d q}{(2\pi)^d}\Delta^4 \tr\left[4im^2\theta\gamma_5 \left(D^2 (2iq\cdot D)^2+2iq\cdot D D^2 2iq\cdot D+(2iq\cdot D)^2 D^3\right)\right]$
    \item At $n=4$: $\int \frac{\dd^d q}{(2\pi)^d}\Delta^5 \tr\left[4im^2\theta\gamma_5 (2iq\cdot D)^4\right]$.
\end{itemize}
In each of these terms, the momenta can be freely commuted to the left for the integration for the same reasons as before. Indeed, if one of the covariant derivatives were localised on a momentum $q$, it would bear no Dirac matrix since $q$ is a Dirac scalar, hence the term would vanish under the trace.
The sum of the different contributions yields,
\begin{align}
2\mathcal{A}^{\mathrm{grav}}=&\tr\,4im^2\theta\gamma_5\bigg(\mathcal{I}[q^0]^3 D^2D^2+\mathcal{I}[q^2]^4g^{\mu\nu}(2i)^2(D^2D_\mu D_\nu+D_\mu D^2D_\nu+D_\mu D_\nu D^2)
\nonumber \\
&\quad\quad\quad\quad\quad+\mathcal{I}[q^4]^5(g^{\mu\nu}g^{\rho\sigma}+g^{\mu\rho}g^{\nu\sigma}+g^{\mu\sigma}g^{\nu\rho})(2i)^4 D_\mu D_\nu D_\rho D_\sigma\bigg)
\nonumber \\
&=\frac{-i}{16\pi^2}\frac{2}{6}\tr\, i\theta \gamma_5\bigg(0D^2D^2+2D^\mu D^\nu D_\mu D_\nu-2D^\mu D^2 D_\mu\bigg)
\nonumber \\
&=\frac{-i}{16\pi^2}\frac{2}{6}\tr\, i\theta\gamma_5 F^{\mu\nu}F_{\mu\nu} \, .
\label{g5FFgrav}
\end{align}
In the last line we have not used any integration by parts nor trace cyclicity, it is only algebra. Note that when computing a gauge anomaly, the contributing term is of the form $\tr\,\gamma_5\sigma.F\sigma.F$, which vanishes in gravity thanks to the use of Eq.~\eqref{eq:sigmaFgrav} earlier (and because we discarded the gauge sector).

Now one must pay some attention to the last line of Eq.~\eqref{g5FFgrav}. The field strength on the right is in the fermion representation, we can thus write,
\begin{equation}
\frac{-i}{16\pi^2}\frac{2}{6}\tr\,i\theta F^{\mu\nu}F^\psi_{\mu\nu}\, .
\end{equation}
However, the field strength on the left will contract the indices of $F^\psi_{\mu\nu}$ because of the Christoffel connection, 
\begin{equation}
F^{\mu\nu}F^\psi_{\mu\nu}=F^{\psi,\mu\nu}F^\psi_{\mu\nu}+\gamma^\alpha\gamma^\beta [ R R ]\, ,
\end{equation}
where the last term is a sum of Riemann tensors contracted together and with the two Dirac matrices. 
It vanishes using the symmetries of the tensors (and also vanishes under the Dirac trace).

Using Eq.~\eqref{eq:Fgrav}, we obtain,
\begin{equation}
{\mathcal A}^{\text{grav}}=\frac{-i}{16\pi^2}\frac{1}{6}\tr\,\left( i\theta\gamma_5\frac{1}{4}\gamma^\alpha\gamma^\beta\tensor{R}{^\mu^\nu_\alpha_\beta}\gamma^\mu\gamma^\nu\frac{1}{4}\tensor{R}{_\rho_\sigma_\mu_\nu}\right)=\frac{-i}{384\pi^2}\bar\epsilon^{\mu\nu\rho\sigma}\tensor{R}{^\alpha^\beta_\mu_\nu}\tensor{R}{_\alpha_\beta_\rho_\sigma} \, ,
\end{equation}
which is the so-called axial-gravitational anomaly. We have $\bar\epsilon^{\mu\nu\rho\sigma}=\epsilon^{\mu\nu\rho\sigma}/\g$.\\

We can notice that in our computation, the only contribution to the gravitational anomaly is in the end the spin-connection via the field strengths, although there are many terms with covariant derivatives that are localised on propagators that can yield Riemann squared terms via the Christoffel connection. 
This translates the fact that a fermion in curved space-time is not subject to diffeomorphism invariance, but only to Lorentz invariance. The spin-connection only ensures that Lorentz invariance is preserved in curved space-time. Therefore, it is expected that one can get the gravitational anomaly only considering the spin-connection, and not minding the Christoffel connection.

In the end, we could have had the correct result in a very simple framework where space-time is considered flat, but the covariant derivatives acting on a spinor bears the spin-connection (the covariant derivative acting on a Dirac matrix would be zero since we would consider it "uncharged" under the spin-connection).

\section{Scale anomaly}

It is well-known that there are two main categories of symmetries which are broken by the quantisation of a theory. The first is the axial symmetry associated with Dirac's $\gamma_5$, the chiral anomaly, that we have just treated in details. The other is the Weyl transformation, which changes the length scale of space-time, keeping the local angle invariant, this is called the Weyl anomaly or conformal or trace or scale anomaly~\cite{Coleman:1970je, Crewther:1972kn, Fujikawa:1980vr, PhysRevD.23.2262, Adler:1976zt}. We then propose to evaluate the Weyl anomaly always following the prescription described in section~\ref{sec:Main} and for pedagogical reasons we stick to the case of QED,
\begin{equation}
\mathcal{L}=\bar\psi(i\slashed\pa-\slashed V-m)\psi-\frac{1}{4e^2}F^2\, .
\end{equation}

Scale invariance is classically broken by the fermion mass term. The divergence of the Noether current $J^\mu$ associated to the scale transformation, i.e the trace of the symmetric energy-momentum tensor $\tilde T_\mu^\mu$, reads,
\begin{equation}
     \partial_\mu J^\mu = \tilde T_\mu^\mu   = m \bar \psi \psi\, .
\end{equation}
This relation is also broken at the quantum level by the renormalisation of the coupling $e$. 

The scale transformation $x_\mu \to x'_\mu= e^\sigma x_\mu$ induces,
\begin{equation}
    \begin{split}
        \frac{\partial}{\partial x^\mu} &\to \frac{\partial}{\partial x'^\mu}=e^{-\sigma}\frac{\partial}{\partial x^\mu} \, ,\\
        \dd^d x &\to \dd^dx'=e^{d\sigma} \dd^d x \, ,
    \end{split}
\end{equation}
and the fields transform with their canonical mass dimension,
\begin{equation}
    \begin{split}
        \psi(x) &\to \psi'(x')= e^{-(d-1)\sigma/2}\psi(x)\, ,\\
        \bar\psi(x) &\to \bar\psi'(x')= e^{-(d-1)\sigma/2}\bar\psi(x)\, ,\\
        A_\mu(x)&\to A'_\mu(x')=A_\mu(x')=e^{-\sigma}A_\mu(x)\, ,
    \end{split}
\end{equation}
where $d$ is the dimension of space-time.
Notice that the gauge field does not transform by itself, it only transforms due to its dependence on $x$ \cite{Fujikawa:2004cx}.

Using the invariance of the path integral under the relabelling of the path integral variables, and the invariance of the space-time integral under relabelling the space-time variable, we can write,
\begin{equation}
\int (\dc\psi)'(\dc\bar\psi)' \exp\left(i\int\dd^dx'\mathcal{L}[x',\psi'(x'),A_\mu(x')]\right)
=\int \dc\psi\dc\bar\psi \exp\left(i\int\dd^dx\mathcal{L}[x,\psi(x),A_\mu(x)]\right) \, .
\end{equation}
On the other hand we know how the action transforms, and we can assume that the transformation of the measure produces a Jacobian,
\begin{align}
\begin{split}
&\int (\dc\psi)'(\dc\bar\psi)' \exp\left(i\int\dd^dx'\mathcal{L}[x',\psi'(x'),A_\mu(x')]\right)\\
&=\int J[\sigma]\dc\psi\dc\bar\psi \exp\left(i\int\dd^dx\mathcal{L}[x,\psi(x),A_\mu(x)]+\int\dd^dx\left(\bar\psi(-i\frac{d-1}{2}(\slashed \partial\sigma)-m\sigma)\psi\right)\right) \, .   
\end{split}
\end{align}
One can take advantage of equating the two path integrals to express the Jacobian as,
\begin{equation}
J[\sigma] = \frac{\det(i\slashed D - m)}{\det(i\slashed D - m - \sigma m - i\frac{d-1}{2} (\slashed \partial \sigma))}\, .
\end{equation}
The term proportional to $(\slashed \partial \sigma)$ requires to be regularised, we use the BMHV scheme of dimensional regularisation~\cite{tHooft:1972tcz,Breitenlohner:1977hr} since the calculation does not involve any $\gamma_5$ matrices. At order $m^0$, the divergent contribution from $-i\frac{d-1}{2}(\slashed\pa \sigma)$ vanishes, only the finite contribution from $\sigma m$ remains and yields the scale anomaly,
\begin{equation}
    \mathcal A_{\rm scale} = \frac{\sigma}{24\pi^2}\tr\, (F_{\mu\nu})^2 \, .
\end{equation}
More details of the calculation are provided in appendix \ref{ScaleApp}.
However, higher order terms (terms of order $1/m^k$, with $k>0$) involve contributions from both $\sigma m$ and $\frac{d-1}{2}(\slashed \partial \sigma)$ which cancel one another.\\

We can now relate the anomaly to the $\beta$ function.
At tree level the coupling $e$ does not transform. It however transforms at one loop level, and by definition of the $\beta$ function we have,
\begin{equation}
e \to e + \sigma \beta(e) \, .
\end{equation}
The following action is invariant under the scale transformation up to the mass term,
\begin{equation}
S=\int\dd^dx \mathcal{L}[x,\psi(x),A_\mu(x)]=\int\dd^d x\left(\bar\psi(i\sD-m)\psi-\frac{1}{4e^2}F^2\right) \, .
\end{equation}
By definition of the $\beta$ function, the gauge sector transforms at one loop like,
\begin{equation}
    -\frac{1}{4e^2} (F_{\mu\nu})^2 \to -\frac{1}{4e^2} (F_{\mu\nu})^2 + \sigma \frac{\beta(e)}{2e^3} (F_{\mu\nu})^2.
\end{equation}
By identification with the term produced at one loop by the Jacobian, we deduce following expression for the one loop $\beta$ function,
\begin{equation}
    \beta(e) = \frac{e^3}{12\pi^2}\, ,
\end{equation}
which corresponds to the well-known QED $\beta$ function.\\

A derivation of the scale anomaly has been proposed by Fujikawa in Ref.~\cite{Fujikawa:1980vr}. This is interesting to point out an important difference between our procedure and Fujikawa's procedure for computing the scale anomaly. In Fujikawa's method, we temper directly with the path integral measure, therefore is it necessary to isolate the field transformation from the space-time transformation. This is achieved by introducing the curvature of space-time and defining a diffeomorphism invariant path integral measure. Because of this redefinition of the fields, they do not transform with their canonical mass dimension anymore. It is even emphasised in \cite{Fujikawa:2004cx} that doing the transformation with their canonical mass does not yield the correct result.

However, in our procedure we can use the invariance of the space-time integral under relabeling the space-time variable, along with transforming the fields with their canonical mass dimension. As we showed it provides the correct result, without having to introduce the curvature of space-time, nor redefining the fields in a diffeomorphism invariant way.

\section{Comparison with Fujikawa's method} \label{sectionFujikawa}


Let us consider the simple case of a vector gauge theory and an axial fermion reparametrisation. The covariant derivative is,
\begin{equation}
D_\mu=\pa_\mu+iV_\mu\, .
\end{equation}
Under the infinitesimal Abelian field reparametrisation,
\begin{align}
\begin{split}
&\psi\to e^{i\theta(x)\gamma_5}\psi
\, , \quad
\bar\psi\to \bar\psi e^{i\theta(x)\gamma_5}\, ,
\label{FermReparamSection6}
\end{split}
\end{align}
the path integral,
\begin{equation}
\int\dc\psi\dc\bar\psi e^{\int\dd^4x\bar\psi(i\sD-m)\psi} \ ,
\end{equation}
produces the following Jacobian,
\begin{equation}
J[\theta]=\frac{\det(i\sD-m)}{\det(i\sD-m-(\slashed\pa\theta)\gamma_5-2im\theta\gamma_5)}\, .
\end{equation}
As emphasised before, the initial theory is ill-defined, leading to an ambiguous Jacobian. A well-known way of dealing with the ambiguity is to bosonise it~\cite{Bertlmann:1996xk} (it enforces the conservation of the vector current as explained in section \ref{DetEigReg}),
\begin{equation}
J[\theta]^2=\frac{\det(\sD^2+m^2)}{\det(\sD^2+m^2+\{i\sD,(\slashed\pa\theta)\gamma_5\}+4im^2\theta\gamma_5)}\, .
\end{equation}
Note also that since the theory is non-chiral we have $(i\sD)^\dagger=i\sD$, hence there is no difference between the bosonisation with $\sD^\dagger\sD$ or $\sD^2$.

As detailed in Eq.~\eqref{logJtoTr} of appendix~\ref{RatioDetExp}, the $log$ of this ratio of determinant can be written as follows,
\begin{equation}
\log J[\theta]=-\Tr\left( \half \frac{\{i\sD,(\slashed\pa\theta)\gamma_5\}}{m^2}\frac{1}{1+\frac{\sD^2}{m^2}}\right)-\Tr\left(2i\theta\gamma_5\frac{1}{1+\frac{\sD^2}{m^2}}\right)\, ,
\label{logJ-CDE}
\end{equation}
where $\Tr$ is the trace over both space-time and internal spaces.

Now let's recall Fujikawa's procedure to compute the anomaly. 
The Dirac operator $i\sD$ is hermitian thus provides a complete and orthonormal set of eigenfunctions $\{\varphi_n\}$ with real eigenvalues $\lambda_n$, such that $i\sD\varphi_n=\lambda_n\varphi_n$. We use this basis to decompose the fermion field, this will enable us to define the path integral measure,
\begin{align}
\begin{split}
&\psi(x)=\sum_n a_n\varphi_n(x)
\, , \quad
\bar\psi(x)=\sum_n \varphi^\dagger_n(x)\bar b_n\, .
\end{split}
\end{align}
The measure is then defined as,
\begin{equation}
    \dc\psi\dc\bar\psi=\prod_n da_nd\bar b_n\, .
\end{equation}
Now the fermion undergoes an axial reparametrisation as in Eq.~\eqref{FermReparamSection6}, the reparametrised field can be decomposed in the eigenbasis as well, with coefficients $a_n'$ and $\bar b_n'$. They are related to $a_n$ and $\bar b_n$ by the transformation matrix $C_{nm}$,
\begin{align}
\begin{split}
&a_n'=\sum_m C_{nm}a_m
\, , \quad
\bar b_m'=\sum_n C_{nm}\bar b_n\, ,
\end{split}
\end{align}
where,
\begin{equation}
C_{nm}=\delta_{nm}+i\int\dd^4x\theta(x)\varphi_n^\dagger(x)\gamma_5\varphi_m(x)\, .
\end{equation}
Now we know that the Grassmann measure transforms with the inverse  determinant of the transformation operators,
\begin{align}
\begin{split}
&\prod_n da_n'=(\det C)^{-1}\prod_n d a_n\\
&\prod_m d\bar b_m'=(\det C)^{-1}\prod_m d \bar b_m\, ,
\end{split}
\end{align}
whence,
\begin{equation}
\dc\psi'\dc\bar\psi'=(\det C)^{-2}\dc\psi\dc\bar\psi\, .
\end{equation}
Finally, using $\det=\exp\Tr\log$ and expanding the $log$ in first order in $\theta$ infinitesimal, the Jacobian reads,
\begin{equation}
\log J[\theta]=-2i\int \dd^4x\theta(x)\sum_n\varphi_n^\dagger(x)\gamma_5\varphi_n(x)\, .
\end{equation}
Since $\{\varphi_n\}$ defines a complete set of operators, it is in fact,
\begin{equation}
\log J[\theta]=-\Tr\,2i\theta\gamma_5\, ,
\end{equation}
where the trace \textit{Tr} is over both internal indices and space-time.

This quantity needs to be regularised, hence Fujikawa introduces a regulator that will work as a cut-off. This regulator needs to preserve the spectrum of the theory. In the simple case of a vector gauge theory, a good choice is,
\begin{equation}
\log J[\theta]=-\lim_{\Lambda\to\infty}\Tr\,2i\theta\gamma_5\,f\left(\frac{\sD^2}{\Lambda^2}\right)\, .
\end{equation}
The function $f$ that was introduced, has to be a smooth function such that $f$ and all its derivatives vanish at infinity, and respect the requirement $f(0)=1$. It can for example be,
\begin{equation}
f(x)=\frac{1}{1+x}\, .
\end{equation}
This leaves us with,
\begin{equation}
\log J[\theta]=-\lim_{\Lambda\to\infty}\Tr\,\left(2i\theta\gamma_5\,\frac{1}{1+\frac{\sD^2}{\Lambda^2}}\right)\, .
\label{logJ-Fuji}
\end{equation}

Now let's compare Eq.~\eqref{logJ-CDE} and Eq.~\eqref{logJ-Fuji}. If one takes the infinite mass limit in Eq.~\eqref{logJ-CDE}, it is clear that one recovers Eq.~\eqref{logJ-Fuji} with $\Lambda$ identified as the physical fermion mass. It thus appears that the procedure presented in this paper (the bosonic CDE) not only amounts to Fujikawa's procedure in the infinite mass limit, but moreover generalises it to a finite and physical mass.\\

In Fujikawa's procedure, the infinite mass limit ensures that the Jacobian is of order $m^0$, while in the bosonic CDE, it is the derivative coupling that plays this role\footnote{In the case of a vector-axial theory, we showed in section \ref{consano} that the analytic continuation $A_\mu\to iA_\mu$ to make $i\sD$ hermitian,  and the bosonisation as in Eq.~\eqref{jacbosonCons}, hence Eq.~\eqref{logJ-CDE} replacing $(\slashed\pa\theta)$ by $(\slashed D\theta)$, is still ambiguous. However, we also noted that the ambiguity was carried by the derivative term only. In Fujikawa's procedure, the derivative term vanishes because of the infinite mass limit, this is why the analytic continuation is sufficient to bring forth the consistent anomaly.}.
Besides, in the bosonic CDE, there is no need to add the regulator by hand. Although, in more complicated cases (when the theory is vector-axial for example), there is still a choice on how to define non-ambiguously the Jacobian, which is in the end making a choice of regulator.

As pointed out in the above, in the bosonic CDE the regulator that naturally appears corresponds to,
\begin{equation}
f(x)=\frac{1}{1+x}\, ,
\end{equation}
as in the treatment of Fujikawa. It is known that this specific regulator amounts to doing a Pauli-Villars regularisation \cite{Bertlmann:1996xk}. It is also known that Pauli-Villars regularisation enforces the conservation of the vector current, while the axial one is anomalous. Therefore, in the infinite mass limit, the bosonic CDE amounts to a Pauli-Villars regularisation, hence conserves the vector current. Now since the anomaly is mass independent, we can expect that with a finite mass, the conservation of the vector current will hold.

Another point that is worth noticing is that in the bosonic CDE, the anomaly always arises in the mass term, even if the mass is finite. Likewise, in a Pauli-Villars regularisation, we know that the anomaly is carried by the mass regulating term (see section 6.2 of Ref.~\cite{Fujikawa:2004cx}).\\

\section{Conclusion}

EFT has always been a pillar in particle physics. Its fundamental reason is to transform a QFT paradigm into a phenomenologically accessible one. In practice, it can be used to ``replace'' (integrating-out)  supposedly directly inaccessible fields by potentially observable distortions (higher dimensional operators effects). Those heavy or weakly coupled fields may or may not be sensible to other symmetries.

If these symmetries are for some reasons anomalous, then the EFT will inherite very specific anomalous operators i.e interactions between the ``light'' fields. In order to compute these interactions, it exists very efficient and well-known master formulas and techniques to apply on the full UV theory, mainly by computing loop Feynman diagrams or by computing the variance of the fermionic measure in the path integral which may involve quite a different toolkit. 

Building successively EFTs from the path integral, i.e integrating out fields successively, offers undeniable advantages. Then, when encountering anomalous symmetries one has to inevitably borrow standard calculations made in substantially different context. This task is not always straightforward.

In this work, we have presented a natural way to build EFTs from the path integral, being able to get the anomalous operators for free and in a self consistent way, i.e without having to rely on any external results and then specific conditions of applicability. Indeed, when building EFTs involving anomalous interactions one may face the choice of deciding in which current this has to go, consequently, one may customise the regularisation of EFTs.

While constructing an EFT from a generic UV theory, this method allows to directly obtain the anomalous interactions. The procedure is simple and based on the well-know fact that an anomalous tranformation means that the path integral measure transforms with a non-trivial Jacobian. Opposed to Fujikawa's prescription which computes directly the transformed measures from accessing the zero modes of Dirac operators, we expressed this Jacobian as a ratio of functional determinants i.e two EFTs. The ``comparison'' of these two EFT's allows us to access straightforwardly the anomalous operators. This is even more remarkable when this method is combined with the CDE techniques. In practice, one only has to perform basic algebra and power counting to access the expected result. This all the more striking that the CDE, extended to curved space-time, provides so straightforwardly the gravitational anomaly.

The presented methodology is even more impressive since it allows one to obtain all various types of anomalies in vector-axial gauge theory (covariant anomaly involving vector or axial symmetries; consistent anomaly), the axial gravitational anomaly and the Weyl anomaly (even if the last one has a quite different physical nature).

These computations have been presented in details and the heart of those computations are (as expected) the procedure used to regularise ill-define functional determinants. A first method consists in fermionic CDE combined with dimensional regularisation and introducing free parameters to keep track of $\gamma_5$ ambiguities subsequently fixed when imposing gauge invariance (covariant anomaly) or integrability conditions (consistent anomaly). We have also investigated another method based on evaluating a ``squared Jacobian'' combined with a CDE.

We believe that this is the first time that a method is proposed to evaluate in a general way the covariant and consistent anomaly from the path integral having then the possibility to choose in which current the anomaly has to stand.

We have also presented an enlightening comparison between our work and the seminal work of Fujikawa which allows one to appreciate how an EFT mass expansion can mimic the educated guess regulators used by Fujikawa. Moreover, it appears that the bosonic CDE extends Fujikawa's procedure by replacing the regulator by a physical and finite mass.

Another interesting comparison with Fujikawa's method is highlighted in the computation of the scale anomaly, where we are able to compute the scale anomaly without having to introduce the curvature of space-time, nor redefining the path integral measure in a diffeomorphism invariant way, which substantially simplifies the procedure.

Furthermore, it appeared recently several public codes (see for example Refs.~\cite{Cohen:2020qvb,Fuentes-Martin:2020udw}) based on analytical and systematic CDE to efficiently build EFTs. They also drastically simplify the so-called {\it matching} step. Unfortunately, models involving QFT anomalies are out of reach of these codes. The results of this work, could be straightforwardly implemented in these or similar codes and then allow to compute anomalous interactions in a self-consistent manner in the path integral formalism. Incorporating theoretical models involving anomalous features, as appearing in many BSM models, would have important phenomenological implications.

\section*{Acknowledgments} 
The authors are grateful to Christopher Smith for useful discussions and initial collaborations in this project. This work is supported by the IN2P3 Master projects ``Axions from Particle Physics to Cosmology'' and BSMGA. The work of B.F. is also supported by the Deutsche Forschungsgemeinschaft under Germany’s Excellence Strategy - EXC 2121 Quantum Universe - 390833306.

\appendix

\section{Master integrals}
\label{Appendix:master-integrals}

\begin{table}[h!]
\centering
\begin{tabular}{|c|ccc|}
\hline
$\tilde{\mathcal{I}}[q^{2n_c}]_i^{n_i}$ & $n_c=0$ & $n_c=1$ & $n_c=2$ \T\B  \\
\hline
$n_i=1$ \T
& $M_i^2 \bigl(1-\logm{M_i^2}\bigr)$ 
& $\frac{M_i^4}{4} \bigl(\frac{3}{2} -\logm{M_i^2}\bigr)$ 
& $\frac{M_i^6}{24} \bigl(\frac{11}{6} -\logm{M_i^2}\bigr)$ \\
$n_i=2$ 
& $-\logm{M_i^2}$ 
& $\frac{M_i^2}{2} \bigl(1 -\logm{M_i^2}\bigr)$ 
& $\frac{M_i^4}{8} \bigl(\frac{3}{2} -\logm{M_i^2}\bigr)$ \\
$n_i=3$ 
& $-\frac{1}{2M_i^2}$ 
& $-\frac{1}{4}\logm{M_i^2}$ 
& $\frac{M_i^2}{8} \bigl(1 -\logm{M_i^2}\bigr)$ \\
$n_i=4$ 
& $\frac{1}{6M_i^4}$ 
& $-\frac{1}{12M_i^2}$
& $-\frac{1}{24}\logm{M_i^2}$\\
$n_i=5$
& $-\frac{1}{12M_i^6}$
& $\frac{1}{48M_i^4}$
& $-\frac{1}{96M_i^2}$ \B \\
\hline
\end{tabular}
\caption{Commonly-used master integrals with degenerate heavy particle masses. $\tilde{\mathcal{I}}=\mathcal{I}/\frac{i}{16\pi^2}$ and the $\frac{2}{\epsilon} -\gamma +\log 4\pi$ contributions are dropped.}
\label{tab:MIheavy}
\end{table}

In this appendix, we discuss the master integrals and tabulate some of
them that are useful in practice. In this paper our results are written in terms of master integrals $\,\mathcal{I}$, defined by

\begin{align}
            \int \dfrac{\dd^dq}{(2\pi)^d} \dfrac{q^{\mu_1}\cdots q^{\mu_{2n_c}}}{(q^2-M_i^2)^{n_i}(q^2-M_j^2)^{n_j}\cdots }
            = g^{\mu_1 \cdots \mu_{2n_c}} \mathcal{I}[q^{2n_c}]^{n_i n_j \cdots}_{i j \cdots }
\end{align}
In the mass degenerate case, the master integrals, $\mathcal{I}[q^{2n_c}]^{n_in_j\dots}_{ij\dots}$, reduce to the form $\,\mathcal{I}[q^{2n_c}]_i^{n_i}$, for which the general expression reads,
\begin{equation}
\mathcal{I}[q^{2n_c}]_i^{n_i} = \frac{i}{16\pi^2} \bigl(-M_i^2\bigr)^{2+n_c-n_i}
\frac{1}{2^{n_c}(n_i-1)!} \frac{\Gamma(\frac{\epsilon}{2}-2-n_c +n_i)}{\Gamma(\frac{\epsilon}{2})} \Bigl(\frac{2}{\epsilon} -\gamma +\log 4\pi-\log\frac{M_i^2}{\mu^2}\Bigr) \,,
\end{equation}
where $d=4-\epsilon$ is the space-time dimension, and $\mu$ is the renormalisation scale. In the $\overline{\rm MS}$ scheme, we replace, $\bigg(\dfrac{2}{\epsilon} -\gamma \, + \, \log 4\pi -\log\dfrac{M_i^2}{\mu^2}\bigg)$ by $\bigg(-\log\dfrac{M_i^2}{\mu^2}\bigg)$ in the final result. We factor out the common prefactor, $\mathcal{I}=\frac{i}{16\pi^2}\tilde{\mathcal{I}}$ and present a table of $\tilde{\mathcal{I}}[q^{2n_c}]_i^{n_i}$ for various $n_c$ and $n_i$, needed in our computations, in Table~\ref{tab:MIheavy}.

In this paper, we only need the integrals in the mass degenerate case, i.e $M_i=M$ for all $i$.

\section{Expansion of a ratio of determinants}
\label{RatioDetExp}

The Jacobians that we have to compute are always of the form,
\begin{equation}
J[\theta]=\frac{\det(A)}{\det(A+f(\theta))}
\label{Jexample}
\end{equation}
where $A$ is some operator, and $f(\theta)$ carries all the $\theta$ dependence.

Using $\log\det=\Tr\log$, it can be expanded at first order in $\theta$ (which is infinitesimal) as follows,
\begin{equation}
\log J[\theta]=\Tr\log A-\Tr\log(A+f(\theta))=-\Tr\log\left(1+\frac{f(\theta)}{A}\right)=-\Tr\frac{f(\theta)}{A}+\mathcal{O}(\theta)\, .
\label{logJtoTr}
\end{equation}
If the Jacobian has been bosonised, $A=-\sP^\dagger\sP+m^2$, whereas if it was kept in the fermionic form, $A=i\sD-m$. $f(\theta)$ also depends on whether the Jacobian has been bosonised or not. To fix the ideas consider the fermionic form, although the reasoning holds for both.

We now explicit the trace over space-time, and use the Fourier transform to make momentum appear. We thus have $i\sD-\slashed q-m=\Delta^{-1}(1-\Delta(-i\sD))$, with the propagator $\Delta=-1/(\slashed q+m)$ . We denote $f_q(\theta)$ the Fourier transform of $f(\theta)$. We then proceed with,
\begin{align}
\begin{split}
\log J[\theta]&=-\int \dd^dx\frac{\dd^dq}{(2\pi)^d}\tr\,\frac{f_q(\theta)}{\Delta^{-1}(1-\Delta(-i\sD))}\\
&=-\int \dd^dx\frac{\dd^dq}{(2\pi)^d}\tr\,f_q(\theta)\sum_{n\geq0}\left(\Delta (-i\sD)\right)^n\Delta\, .
\end{split}
\end{align}
It is apparent that the ratio of the two determinants in Eq.~\eqref{Jexample} is proportional to $\theta$.
Now, using the cyclicity of the trace, this determinant can be written under the form,
\begin{align}
\begin{split}
\log J[\theta]&=\int \dd^dx\frac{\dd^dq}{(2\pi)^d}\tr\,\sum_{n\geq0}\left(\Delta (-i\sD)\right)^n \big(\Delta (-f_q(\theta))\big)\\
&=\left.\int \dd^dx\frac{\dd^dq}{(2\pi)^d}\tr\,\sum_{n\geq0}\frac{1}{n+1}\Big[\Delta \big(-i\sD-f_q(\theta)\big)\Big]^{n+1}\right|_{\mathcal{O}(\theta)}\\
&=\left.\int \dd^dx\frac{\dd^dq}{(2\pi)^d}\tr\,\sum_{n\geq1}\frac{1}{n}\Big[\Delta \big(-i\sD-f_q(\theta)\big)\Big]^n\right|_{\mathcal{O}(\theta)}
\end{split}
\end{align}
The factor $1/(n+1)$ appears in the second-to-last line to avoid overcounting the number of terms with one $f_q(\theta)$ insertion.

Note that the use of trace cyclicity despite the presence of $\gamma_5$ is not an issue, since we regularise using the free parameters, or using a proper hermitian operator that makes the Jacobian unambiguous ( $\sD^\dagger\sD$ ).


\section{Fermionic expansion with free parameters: case of anomalous vector symmetry}\label{appfreeparam} 

In this appendix, we recast in details the calculation that is presented in section~\ref{subsec:vector-global-anomalous}. A similar approach have been studied in a different context in Ref.~\cite{Quevillon:2021sfz}. As derived in section \ref{subsec:vector-global-anomalous}, we state here the Jacobian produced by the vector transformation,
\begin{equation}
	J[\theta] = \dfrac{\det \big(i \slashed D -m\big)}{\det\big(i\slashed D -m - (\slashed D \theta)\big)} \ .
	\label{appdix:Jac-vector-rotation}
\end{equation}
Since we introduced the auxiliary background field $\xi_{\mu}$, and the longitudinal mode $\pi_A$ of the axial gauge field $A_{\mu}$, the expansion of the Jacobian now reads,
\begin{equation}
	\tilde{\mathcal A} = \left.\int \frac{\dd^d q}{(2\pi)^d} \sum_{n=1}^\infty \frac{1}{n} \tr \left[ \frac{-1}{\slashed q +m} \left(- i\slashed D + [\slashed \xi -(\slashed \partial \theta)] + m\dfrac{\pi_A}{v}i\gamma_5 \right) \right]^n \right|_{\mathcal{O}(\theta,\,\pi_A)} 
	\, ,
	\label{appdix:atildevec}
\end{equation}
where we restrict our self in the Abelian gauge fields and Abelian $\theta$.

\paragraph{Evaluating $\mathcal{A}_{\slashed \partial}.$} Since the mass term does not explicitly break the vector transformation, only the term involving $(\slashed\pa\theta)$ will contribute to the anomaly as can be seen in Eq.~\eqref{appdix:Jac-vector-rotation} (with $\theta$ Abelian). 
\\

\textbf{At $\mathbf{n=4}$}, we obtain the $m^0$ term,
    \begin{align}
    \tilde{\mathcal{A}}_{\slashed\partial} 
    &= \bigg[ -8m^4\Integral_i^4 + 32m^2\Integral[q^2]_i^4 + \alpha\big(48\varepsilon \big)\Integral[q^4]_i^4  \bigg] \epsilon^{\mu\nu\rho\sigma} \tr \big[\xi_{\mu}-(\partial_{\mu}\theta)\big] (iA_{\nu})\big[i\partial_{\rho}V_{\sigma} \big]
    \nonumber \\
    &= \dfrac{i}{8\pi^2}\big( -1 + \alpha^{\prime} \big)\epsilon^{\mu\nu\rho\sigma} \tr  \big[\xi_{\mu}-(\partial_{\mu}\theta)\big] (iA_{\nu})\big[i\partial_{\rho}V_{\sigma} \big]
    \nonumber \\
    &= \beta^{\prime}\,\epsilon^{\mu\nu\rho\sigma} \tr \, \big[\xi_{\mu} - (\partial_{\mu}\theta) \big] (iA_{\nu}) F^V_{\rho\sigma}
    \, ,
    \label{appdixC: n=4 terms}
    \end{align}
    where we define $\beta^{\prime} = i(-1+\alpha^{\prime})/(8\pi^2)$. In this computation, we followed the strategy outlined in section \ref{sec: ambiguitites-free-parameters} to deal with the ambiguous traces. In $d=4-\epsilon$ dimensions, we explicitly have,
    \begin{align}
    \tr\, g^{abcd}\big( \gamma_a\gamma^{\mu}\gamma_b\gamma^{\nu}\gamma_c\gamma^{\rho}\gamma_d\gamma^{\sigma}\gamma_5 \big) 
    &\rightarrow 
    \alpha_1 \tr\, g^{abcd}\big( \gamma_a\gamma^{\mu}\gamma_b\gamma^{\nu}\gamma_c\gamma^{\rho}\gamma_d\gamma^{\sigma}\gamma_5 \big) + \alpha_2 \tr\, g^{abcd}\big( \gamma_a\gamma^{\mu}\gamma_b\gamma^{\nu}\gamma_c\gamma^{\rho}\gamma_5\gamma_d\gamma^{\sigma} \big) 
    \nonumber \\
    &+ \alpha_3 \tr\, g^{abcd}\big( \gamma_a\gamma^{\mu}\gamma_b\gamma^{\nu}\gamma_5\gamma_c\gamma^{\rho}\gamma_d\gamma^{\sigma} \big) + \alpha_4 \tr\, g^{abcd}\big( \gamma_a\gamma^{\mu}\gamma_5\gamma_b\gamma^{\nu}\gamma_c\gamma^{\rho}\gamma_d\gamma^{\sigma} \big)
    \nonumber \\
    &= \varepsilon\big(\alpha_1 -\alpha_2 + \alpha_3 -\alpha_4 \big) \big[ i 24 \, \epsilon^{\mu\nu\rho\sigma} \big]
    \nonumber \\
    &= \varepsilon \alpha^{\prime} \big[ i 24 \epsilon^{\mu\nu\rho\sigma} \big]
    \, .
    \label{appdix:axial-type1-trace}
\end{align}
    Even though we enforced the condition $\alpha_1 + \alpha_2 + \alpha_3 +\alpha_4 =1$, the above trace still depends on free parameters and thus is ambiguous. In the last line of Eq.~\eqref{appdix:axial-type1-trace}, we relabeled the sum of free parameters by a new parameter, $\alpha^{\prime}$. Note that the standard evaluation of BMHV's scheme without free parameters is equivalent to $\alpha^{\prime} = 1$.
    
    As a remark, Eq.~\eqref{appdixC: n=4 terms} is not gauge invariant due to the presence of the Chern-Simons term, $\epsilon^{\mu\nu\rho\sigma}\,\tr\,\xi_{\mu}(iA_{\nu})F^V_{\rho\sigma}$\,.
    \\

\textbf{At $\mathbf{n=5}$}, the order $m^0$ terms related to the Goldstone boson $\pi_A$ and the auxiliary field $\xi_{\mu}$ that we need to enforce the gauge invariance of the axial current are,
\begin{align}
    \tilde{\mathcal{A}}_{\slashed\partial} &= -\dfrac{i}{8\pi^2} \epsilon^{\mu\nu\rho\sigma} \tr \bigg[ \dfrac{\pi_A}{v} (\partial_{\mu}\xi_{\nu})F^V_{\rho\sigma} \bigg]\, .
\end{align}

\paragraph{Enforcing gauge invariance.}
In the Abelian case, a complete set of gauge transformations is
\begin{align}
\begin{cases}
    V_{\mu} \rightarrow V_{\mu} + (\partial_{\mu}\varepsilon_V)
    \, ,
    \\ 
    A_{\mu} \rightarrow A_{\mu} + (\partial_{\mu}\varepsilon_A)
    \, ,
    \\
    \pi_A \rightarrow \pi_A -2v\varepsilon_A
    \, ,
    \label{appdix:local-GT}
\end{cases}
\end{align}
where $\varepsilon_{V,A}$ are infinitesimal gauge parameters. Under the gauge transformation of Eq.~\eqref{appdix:local-GT}, we enforce,
\begin{align}
\delta_G\big(\tilde{\mathcal{A}}_{\slashed\partial}\big)
    &=
    \delta_G \, \epsilon^{\mu\nu\rho\sigma}\bigg[ 
    \beta\, \tr \, \big[\xi_{\mu} - (\partial_{\mu}\theta) \big] (iA_{\nu}) F^V_{\rho\sigma}
    - \dfrac{i}{8\pi^2} \, \tr\, \dfrac{\pi_A}{v} (\partial_{\mu}\xi_{\nu})F^V_{\rho\sigma} \,
    \bigg] = 0
    \, .
\end{align}
Hence, we are now able to fix the value of the free parameter,
\begin{align}
    \beta^{\prime} = \dfrac{-i}{8\pi^2}
    \, .
\end{align}
As a small remark in parallel with the usual Feynman diagrams technique, the gauge invariant combination of the General Chern-Simons term, $\epsilon^{\mu\nu\rho\sigma}\tr\big[\xi_{\mu}-(\partial_{\mu}\theta)\big](iA_{\nu})F^V_{\rho\sigma}$, and the Goldstone term, $\epsilon^{\mu\nu\rho\sigma}\tr\,(\pi_A/v) (\partial_{\mu}\xi_{\nu})F^V_{\rho\sigma}$, is equivalent to enforcing the classical Ward identity of the axial current in the massive case.
\\

Eventually, we substitute the value of $\beta^{\prime}$ into Eq.~\eqref{appdixC: n=4 terms}, we then set $\xi_{\mu} \rightarrow 0$, and perform integration by parts. Going back to non-Abelian gauge fields and $\theta$, we obtain the non-Abelian covariant anomaly in the vector current,
\begin{align}
    \mathcal{A} = \mathcal{A}_{\slashed\partial}
    = \dfrac{-i}{16\pi^2}\epsilon^{\mu\nu\rho\sigma} \tr \, \theta \big( F_{\mu\nu}^V F^A_{\rho\sigma} + F_{\mu\nu}^A F^V_{\rho\sigma} \big)\, .
\end{align}

\section{Covariant anomaly: bosonised form} \label{appbosonized}

In this appendix, we detail the bosonisation and the computation of the covariant anomaly in the bosonised form as discussed in sections~\ref{covbosform} and ~\ref{covbosformvec} . For more details about bosonisation, we refer the reader to Refs.~\cite{Ball:1988xg,Bertlmann:1996xk}.

\subsection*{Bosonisation}

The operator $\sP=i\sD$ is not hermitian,
\begin{equation}
\sP^\dagger=(i\slashed\pa-\slashed V-\slashed A\gamma_5)^\dagger=i\slashed\pa-\slashed V+\slashed A\gamma_5,
\end{equation}
therefore, it does not have a well defined eigenvalue problem. However, $\sP^\dagger\sP$ and $\sP\sP^\dagger$ are hermitian, hence they admit two orthogonal eigenbasis with real eigenvalues,
\begin{equation}
	\sP^\dagger\sP \phi_n = \lambda_n^2 \phi_n,\quad \sP\sP^\dagger\varphi_n= \lambda_n^2 \varphi_n, \quad n\in \mathbb{N}, \quad \lambda_n\in \mathbb{R}.
\end{equation}
where
\begin{equation}
\sP\phi_n=\lambda_n\varphi_n\,\text{, }\,\sP^\dagger\varphi_n=\lambda_n\phi_n\,\text{ with }\,\lambda_n\in\mathds{R} \ .
\end{equation}
By decomposing $\psi$ on the orthonormal basis $\{\phi_n\}$, and $\bar\psi$ on the orthonormal basis $\{\varphi_n\}$, we can show that,
\begin{equation}
\det(\sP-m)=\int\dc\psi\dc\bar\psi e^{\int\dd^4x\bar\psi(\sP-m)\psi}=\mathcal{N}\Pi_n(\lambda_n-m),
\end{equation}
and similarly,
\begin{equation}
\det(-\sP^\dagger-m)=\int\dc\psi\dc\bar\psi e^{\int\dd^4x\bar\psi(-\sP^\dagger-m)\psi}=\mathcal{M}\Pi_n(-\lambda_n-m) \ .
\end{equation}
$\mathcal{N}$ and $\mathcal{M}$ depend on the determinant of the matrices that relate $\psi$ to $\phi_n$ and $\bar\psi$ to $\varphi_n$.
They play no role in the computation of the anomaly \cite{Bertlmann:1996xk}.

We can therefore conclude that,
\begin{align}
\begin{split}
|\det(\sP-m)|^2&=\det(\sP-m)\det(\sP^\dagger-m)\\
&=\det(\sP-m)\det(-\sP^\dagger-m)\\
&=\mathcal{N}\mathcal{M}\Pi_n(\lambda_n-m)(-\lambda_n-m)\\
&=\mathcal{N}\mathcal{M}\Pi_n(-\lambda_n^2+m^2)\\
&=\mathcal{N}\mathcal{M}\det(-\sP^\dagger \sP+m^2)=\mathcal{N}\mathcal{M}\det(-\sP\sP^\dagger+m^2) \ .
\end{split}
\end{align}
In the second line, we have used the fact that the determinant is invariant under the change of sign of the Dirac matrices. This can be understood by writing the determinant as a trace using $\log\det=\Tr\log$, and the fact that a trace of an odd number of Dirac matrices vanishes, hence allowing us to flip their sign. Besides, note that under this sign flip, $\gamma_5$ does not change since it's composed of an even number of Dirac matrices.

The sign flip of the Dirac matrices has two purposes. Firstly, it rids us of the cross term between $m$ and $\sP$. Indeed, without the change of sign, we would have,
\begin{equation}
\det(\sP^\dagger-m)\det(\sP-m)=\mathcal{N}\mathcal{M}\det(\sP^\dagger\sP+m^2-m\sP-m\sP^\dagger)\, .
\end{equation}
Secondly, after the Fourier transform, it provides the good relative sign between the $q^2$ and the $m^2$ terms, which allows us to factorise the propagator $\Delta=1/(q^2-m^2)$ instead of $\Delta=1/(q^2+m^2)$ without sign flip.

Now, for a determinant of the form,
\begin{equation}
\det(\sP-m+A),
\end{equation}
where $A$ is a non-diagonal matrix in Dirac and in gauge space, we cannot easily write the determinant in terms of eigenvalues of the Dirac operator, because $A$ is non-diagonal. Therefore, we use naively the product of determinants\footnote{In general the group homomorphism property: $\det(A.B)=\det(A)\det(B)$ is not correct for regularised determinants. However, the determinants we temper with are not regularised yet. Nonetheless, it is generally accepted that for non-regularised determinants one has $\log\det A=\Tr\log A$, which trivially implies the group homomorphism property of the determinant on non-regularised matrices. } to get,
\begin{align}
&\det\left[ e^{i\theta\gamma_5}(\sP^\dagger-m)e^{i\theta\gamma_5}\right]\det\left[e^{i\theta\gamma_5}(-\sP-m)e^{i\theta\gamma_5}\right]
=\mathcal{N}\mathcal{M}\det \left[-\sP^\dagger \sP+m^2+m(\sP-\sP^\dagger)+f(\theta)\right],
\end{align}
where
\begin{align}
\begin{split}
f(\theta)=&4im^2\theta\gamma_5-i[\theta,P^2]\gamma_5-\half[\sigma.F^V,\theta]\gamma_5-\half[\sigma.F^A\gamma_5,\theta]\gamma_5\\
&+2im\left(\theta\gamma_5\sP-\sP\theta\gamma_5\right)+im\left((\sP\theta)-(\sP^\dagger\theta)\right)\gamma_5 \ .
\end{split}
\end{align}
We have used the convenient formulae,
\begin{equation}
\sP^\dagger\sP=\slashed D^{\dagger}\slashed D=-P^2+\frac{i}{2}\sigma^{\mu\nu}[D_\mu,D_\nu]
\end{equation}
and
\begin{equation}
[D_\mu,D_\nu]=F^V_{\mu\nu}+F^A_{\mu\nu}\gamma_5 \ ,
\end{equation}
with the Bardeen curvatures as defined in Eqs.~\eqref{eq:FV} and \eqref{eq:FA}.
As a result we get the Jacobian in the bosonised form,
\begin{equation}
J^2[\theta]=\frac{\det (-\sP^\dagger \sP+m^2)}{\det (-\sP^\dagger\sP+m^2+m(\sP-\sP^\dagger)+f(\theta))}\ .
\end{equation}
Expanding this ratio of determinants we obtain,
\begin{equation}
	2 \mathcal A = \int \frac{\dd^d q}{(2\pi)^d} e^{iqx} \tr \left(f(\theta)+m(\sP-\sP^\dagger)\right)\frac{1}{-\sP^\dagger \sP+m^2} e^{-iqx} \ .
\end{equation}
At this point, one can notice that the term $-m(\sP+\sP^\dagger)$, and the term $-2im\left(\sP^\dagger\theta-\theta i\sD\right)\gamma_5$ in $f(\theta)$ can be dropped since they produce terms with an odd number of Dirac matrices, which will vanish under the Dirac trace.

Finally, let's compute the Fourier transform of $\sP^\dagger\sP$,
\begin{align}
e^{-iqx}\sP^\dagger\sP e^{iqx}&=\left(e^{-iqx}\sP e^{iqx}\right)^\dagger\left(e^{-iqx}\sP e^{iqx}\right)
=\sP^\dagger\sP-\sP^\dagger\slashed q-\slashed q\sP +\slashed q^2
\end{align}
Because of the presence of the axial field, $P_\mu$ does not commute with the Dirac matrices. In order to proceed with the computation let's define,
\begin{equation}
P_\mu=P^V_\mu-A_\mu\gamma_5 \,\text{ where }\, P^V_\mu=i\pa_\mu-V_\mu \ ,
\end{equation}
with $(\sP^V)^\dagger=\sP^V$ and $(\slashed A\gamma_5)^\dagger=-\slashed A\gamma_5$.

Therefore we can write,
\begin{equation}
-\sP^\dagger\slashed q-\slashed q\sP=-\{\sP^V,\slashed q\}+\slashed A\gamma_5\slashed q-\slashed q\slashed A\gamma_5=-2q\cdot P^V-2\{
\slashed A,\slashed q\}\gamma_5=-2q\cdot P^V-2q.A\gamma_5=-2q\cdot P
\label{DdagDFourier1}
\end{equation}
Finally we have,
\begin{equation}
e^{-iqx}\sP^\dagger\sP e^{iqx}=\sP^\dagger\sP-2q\cdot P+q^2 \ .
\label{DdagDFourier2}
\end{equation}

\subsection*{Computation}

We expand the Jacobian using the mass expansion:
\begin{equation}
2 \mathcal{A} = \int \frac{\dd^d q}{(2\pi)^d} \tr\, h(\theta)\sum_{n\geq0} \left[\Delta\left(-P^2+\frac{i}{2}\sigma.F^V+\frac{i}{2}\sigma.F^A\gamma_5+2q\cdot P\right)\right]^n\Delta\, ,
\end{equation}
where
\begin{equation}
h(\theta)=-i[\theta,(P-q)^2]-\half[\sigma.F^V,\theta]\gamma_5-\half[\sigma.F^A\gamma_5,\theta]\gamma_5+4im^2\theta\gamma_5 \ ,
\end{equation}
and $\Delta=\frac{1}{q^2-m^2}$.

Now we gather the terms of order $m^0$, and that have an odd number of $\gamma_5$.
Firstly, consider the contributions from the term $-i[\theta,(P-q)^2]$,
\begin{align}
\int\frac{\dd^dq}{(2\pi)^4}\tr\,(-i[\theta,(P-q)^2])\bigg( &\Delta^2(-P^2+\frac{i}{2}\sigma.F^V+\frac{i}{2}\sigma.F^A\gamma_5) + \Delta^3 (2q\cdot P)^2\bigg)\, .
\end{align}
These contributions vanish under the Dirac trace by lack of Dirac matrices.
Secondly the contributions from the terms $-\half[\sigma.F^V,\theta]\gamma_5-\half[\sigma.F^A\gamma_5,\theta]\gamma_5$,
\begin{align}
\label{sigF1}
\int\frac{\dd^dq}{(2\pi)^4}\tr\, (-\half[\sigma.F^V,\theta]\gamma_5-\half[\sigma.F^A\gamma_5,\theta]\gamma_5)\bigg(&\Delta^2(-P^2)\\
\label{sigF2}
+&\Delta^2(\frac{i}{2}\sigma.F^V+\frac{i}{2}\sigma.F^A\gamma_5)\\
\label{sigF3}
+&\Delta^3(2q\cdot P)^2\bigg)\, .
\end{align}
The terms from Eq.~\eqref{sigF1} and \eqref{sigF3} both vanish under the Dirac trace by lack of Dirac matrices. The term in Eq.~\eqref{sigF2} vanishes too using trace cyclicity in gauge space (all the operators are local).
In the last two contributions, some of the integrals are divergent, but we do not need to compute them to see that the terms vanish, it is the operator itself that vanishes. Therefore, no ambiguity related to the $\gamma_5$ in dimensional regularisation can arise.
Finally, we consider the contributions from the mass term $4im^2\theta\gamma_5$. Note that for this term, all the integrals are finite.
\begin{align}
\label{mterm1}
&\int\frac{\dd^4q}{(2\pi)^4}\tr\,4im^2\theta\gamma_5\Bigg[\Delta^5(2q\cdot P)^4+\Delta^3(-P^2)^2+\Delta^3(\frac{i}{2}\sigma.F^V+\frac{i}{2}\sigma.F^A\gamma_5)^2\\
\label{mterm2}
&+\Delta^4\left( (\frac{i}{2}\sigma.F^V+\frac{i}{2}\sigma.F^A\gamma_5)(2q\cdot P)^2+2q\cdot P(\frac{i}{2}\sigma.F^V+\frac{i}{2}\sigma.F^A\gamma_5)2q\cdot P+(2q\cdot P)^2(\frac{i}{2}\sigma.F^V+\frac{i}{2}\sigma.F^A\gamma_5) \right)\Bigg]\, .
\end{align}
The first two terms from Eq.~\eqref{mterm1} vanish under the Dirac trace. The third term from Eq.~\eqref{mterm1} does contribute and actually yields the covariant anomaly. The term from Eq.~\eqref{mterm2} vanishes under the Dirac trace, by lack of Dirac matrices.
Therefore, among all the possible combinations, in the end only one term contribute:
\begin{align}
\begin{split}
2\mathcal{A}&=\int\frac{\dd^d q}{(2\pi)^d}\Delta^3\tr\,4im^2\theta\gamma_5(\frac{i}{2}\sigma.F^V+\frac{i}{2}\sigma.F^A\gamma_5)^2\\
&=\frac{-i}{16\pi^2}\frac{1}{2m^2}\tr\,4im^2\theta\gamma_5\left((\frac{i}{2}\sigma.F^V)^2+(\frac{i}{2}\sigma.F^A\gamma_5)^2\right)\\
&=\frac{-i}{16\pi^2}\frac{1}{8}\tr\,i\theta\gamma_5\left([\gamma^\mu,\gamma^\nu][\gamma^\rho,\gamma^\sigma]F^V_{\mu\nu}F^V_{\rho\sigma}+[\gamma^\mu,\gamma^\nu]\gamma_5[\gamma^\rho,\gamma^\sigma]\gamma_5 F^A_{\mu\nu}F^A_{\rho\sigma} \right)\\
&=2\frac{-i}{16\pi^2}\epsilon^{\mu\nu\rho\sigma}\tr\,\theta(F^V_{\mu\nu}F^V_{\rho\sigma}+F^A_{\mu\nu}F^A_{\rho\sigma})
\end{split}
\end{align}
where we have discarded terms with even number of $\gamma_5$ matrices (they cannot yield a boundary term so cannot contribute to the final result). The remaining trace in the last line is the trace over the gauge space.

\section{Scale anomaly}
\label{ScaleApp}

In this appendix, we detail the computation of the following Jacobian,
\begin{equation}
J[\sigma] = \frac{\det(i\slashed D - m)}{\det(i\slashed D - m - \sigma m - i\frac{d-1}{2} (\slashed \partial \sigma))}\, ,
\end{equation}
where $d$ is the dimension of space-time, and $\sigma$ is a local scalar function. No $\gamma_5$ is involved in the computation therefore the Jacobian is well-defined, and the computation in $d$ dimensions is performed using BMHV scheme.
This Jacobian can be expanded following the usual procedure described in this paper,
\begin{equation}
\mathcal{A}=-\int\frac{\dd^dq}{(2\pi)^d}\tr\,\left(-i\frac{d-1}{2}(\slashed\pa\sigma)-\sigma m\right)\sum_{n\geq0}\left[\Delta(-i\sD)\right]^n \Delta\, .
\end{equation}
$\Delta=-1/(\slashed q+m)=\Delta_f+\Delta_b$, where we label fermionic propagator $\Delta_f=-\slashed q/(q^2+m^2)$, and bosonic propagator $\Delta_b=m/(q^2-m^2)$.
We focus on the terms of order $m^0$ only. The higher order terms see a cancellation between the mass term and the derivative term.
\paragraph{Derivative term $-i\frac{d-1}{2}(\slashed\pa\sigma)$.}
The only term of order $m^0$ is,
\begin{equation}
\int\frac{\dd^dq}{(2\pi)^d}\tr\,\left(-i\frac{d-1}{2}(\slashed\pa\sigma)\right)\left(\Delta(-i\sD)\right)^3\Delta=\int\frac{\dd^dq}{(2\pi)^d}\tr\,\left(\frac{d-1}{2}(\slashed\pa\sigma)\right)\left(\Delta\sD\right)^3\Delta\, .
\end{equation}
The propagators $\Delta$ now have to be decomposed in terms of fermionic and bosonic propagators $\Delta=\Delta_f+\Delta_b$, keeping in mind that the integral over momentum vanishes if the integrand bears an odd power in momentum. That is to say, the non-vanishing terms have an even number of fermionic propagators. 

Note that there is a factor $\frac{i}{16\pi^2}$ from the master integrals that we discard for now for clarity. It will be accounted for at the end.
\begin{itemize}
    \item The term with only bosonic propagators is,
\begin{align}
\frac{d-1}{2}m^4\mathcal{I}[q^0]^4\tr\,\left( (\slashed\pa\sigma)\sD^3 \right)
=\frac{1}{4}\left(4g^{\mu\sigma}g^{\nu\rho}-4g^{\mu\rho}g^{\nu\sigma}+4g^{\mu\nu}g^{\rho\sigma}  \right)\tr\,(\pa_\mu\sigma)D_\nu D_\rho D_\sigma\, .
\end{align}
    \item For the term with two fermionic propagators, we have to account for all the possible positions (2 fermionic propagators among 4 propagators, thus 6 possibilities),
\begin{align}
&\frac{d-1}{2}m^2\mathcal{I}[q^2]^4g_{ab}\tr\,\Bigg( (\slashed\pa\sigma) \bigg( \gamma^a\sD\gamma^b\sD^2+\gamma^a\sD^2\gamma^b\sD+\gamma^a\sD^3\gamma^b\nonumber\\
&\quad\quad\quad\quad\quad+\sD\gamma^a\sD\gamma^b\sD+\sD\gamma^a\sD^2\gamma^b+\sD^2\gamma^a\sD\gamma^b \bigg)  \Bigg)\nonumber\\
&=-\frac{1}{8} \left( -16g^{\mu\sigma}g^{\nu\rho}+32g^{\mu\rho}g^{\nu\sigma}-16g^{\mu\nu}g^{\rho\sigma} \right)\tr\,(\pa_\mu\sigma)D_\nu D_\rho D_\sigma\, ,
\end{align}
    \item Finally, there is a term with four fermionic propagator insertions. This is the only term that is divergent and that is computed in $d=4-\epsilon$ dimensions,
\begin{align}
&\frac{d-1}{2}\mathcal{I}[q^4]^4g_{abcd}\tr\,\left( (\slashed\pa\sigma)\gamma^a\sD\gamma^b\sD\gamma^c\sD\gamma^d \right)\\
&=\bigg( -\frac{13}{3}g^{\mu\sigma}g^{\nu\rho}+\frac{23}{3}g^{\mu\rho}g^{\nu\sigma}-\frac{13}{3}g^{\mu\nu}g^{\rho\sigma}\\
&\quad\quad+\log\left(\frac{m^2}{\mu^2}\right)\left( -2g^{\mu\sigma}g^{\nu\rho}+4g^{\mu\rho}g^{\nu\sigma}-2g^{\mu\nu}g^{\rho\sigma} \right) \bigg)\tr\,(\pa_\mu\sigma)D_\nu D_\rho D_\sigma \, ,\nonumber
\end{align}
where $g_{abcd}=g_{ab}g_{cd}+g_{ac}g_{bd}+g_{ad}g_{bc}$.
\end{itemize}
Those three contributions come together to yield,
\begin{align}
&\left( -\frac{4}{3}g^{\mu\sigma}g^{\nu\rho}+\frac{8}{3}g^{\mu\rho}g^{\nu\sigma}-\frac{4}{3}g^{\mu\nu}g^{\rho\sigma}+\log\left(\frac{m^2}{\mu^2}\right)\left( -2g^{\mu\sigma}g^{\nu\rho}+4g^{\mu\rho}g^{\nu\sigma}-2g^{\mu\nu}g^{\rho\sigma} \right) \right)\tr\,(\pa_\mu\sigma)D_\nu D_\rho D_\sigma\nonumber\\
&=\left( -g^{\mu\sigma}g^{\nu\rho}+2g^{\mu\rho}g^{\nu\sigma}-g^{\mu\nu}g^{\rho\sigma} \right)\left( \frac{4}{3}+2\log\left(\frac{m^2}{\mu^2}\right) \right)\tr\,(\pa_\mu\sigma)D_\nu D_\rho D_\sigma\, .
\label{ScaleDerTerm}
\end{align}
Now let's explicit the operator,
\begin{equation}
\tr\,(\pa_\mu\sigma)D_\nu D_\rho D_\sigma=-\tr\,\sigma\pa_\mu D_\nu D_\rho D_\sigma\, ,
\end{equation}
up to a boundary term. Writing explicitly $D_\mu=\pa_\mu+V_\mu$, and distributing the derivatives we can write\footnote{Recall that a partial derivative to the right vanishes.},
\begin{equation}
-\tr\,\sigma\pa_\mu D_\nu D_\rho D_\sigma=-\sigma\tr\,\Big[(\pa_{\mu\nu\rho}V_\sigma)+(\pa_\mu V_\nu)(\pa_\rho V_\sigma)+V_\nu(\pa_{\mu\rho}V_\sigma)+(\pa_{\mu\nu}V_\rho V_\sigma)+(\pa_\mu V_\nu V_\rho V_\sigma)\Big]\, .\,
\label{ScaleOpExp}
\end{equation}
where the parenthesis on the derivatives means that it acts locally in everything inside the parenthesis.
It is then simple algebra to show that,
\begin{equation}
(-g^{\mu\sigma}g^{\nu\rho}+2g^{\mu\rho}g^{\nu\sigma}-g^{\mu\nu}g^{\rho\sigma})\tr\,\pa_\mu D_\nu D_\rho D_\sigma=0\, ,
\end{equation}
using Eq.~\eqref{ScaleOpExp} and trace cyclicity. As a result Eq.~\eqref{ScaleDerTerm} vanishes, and the derivative coupling has no contribution at order $m^0$.

\paragraph{Mass term $\sigma m$.}
The only term of order $m^0$ is,
\begin{equation}
-\int\frac{\dd^dq}{(2\pi)^d}\tr\,(-\sigma m)\left(\Delta(-i\sD)\right)^4\Delta=\int\frac{\dd^dq}{(2\pi)^d}\sigma m\tr\,\left(\Delta\sD\right)^4\Delta\, .
\end{equation}
Once again, the only non-vanishing terms are those that involve an even number of fermionic propagators.
The term with only bosonic propagators is,
\begin{align}
&\sigma m^6\mathcal{I}[q^0]^5\tr\,\sD^4
=-\sigma\frac{1}{12}\left(4g^{\mu\sigma}g^{\nu\rho}-4g^{\mu\rho}g^{\nu\sigma}+4g^{\mu\nu}g^{\rho\sigma}  \right)\tr\,D_\mu D_\nu D_\rho D_\sigma\, .
\end{align}
The term with two fermionic propagators has $\binom{5}{2}=10$ contributions. We do not write them all for clarity,
\begin{align}
&\sigma m^4\mathcal{I}[q^2]^5g_{ab}\tr\,\left( \gamma^a\gamma^\mu\gamma^b\gamma^\nu\gamma^\rho\gamma^\sigma+ \gamma^a\gamma^\mu\gamma^\nu\gamma^b\gamma^\rho\gamma^\sigma+... \right)\tr\,D_\mu D_\nu D_\rho D_\sigma
\nonumber \\
&=\sigma \frac{1}{48}\left( 32g^{\mu\rho}g^{\nu\sigma}-16g^{\mu\sigma}g^{\nu\rho} \right)\tr\,D_\mu D_\nu D_\rho D_\sigma\, .
\end{align}
Finally, the term with four fermionic propagators has $\binom{5}{4}=5$ contributions. Again, we do not write them all for clarity,
\begin{align}
&\sigma m^2\mathcal{I}[q^4]^5g_{abcd} \, \tr\,\left( \gamma^a\gamma^\mu\gamma^b\gamma^\nu\gamma^c\gamma^\rho\gamma^d\gamma^\sigma+\gamma^a\gamma^\mu\gamma^b\gamma^\nu\gamma^c\gamma^\rho\gamma^\sigma\gamma^d+... \right)D_\mu D_\nu D_\rho D_\sigma\nonumber\\
&=-\sigma\frac{1}{96}\left( 64g^{\mu\sigma}g^{\nu\rho}-32g^{\mu\rho}g^{\nu\sigma}-32g^{\mu\nu}g^{\rho\sigma} \right)\tr\,D_\mu D_\nu D_\rho D_\sigma\, .
\end{align}
Note that the three contributions are finite, hence the computation is performed in 4 dimensions.
Putting together the three contributions, we obtain,
\begin{equation}
\sigma\tr\,\left( -\frac{4}{3}D^\mu D^2 D_\mu+\frac{4}{3}D^\mu D^\nu D_\mu D_\nu +0D^2D^2 \right)=\sigma\frac{2}{3}\tr\,F^{\mu\nu}F_{\mu\nu}\, .
\end{equation}
\paragraph{Final result.}
Finally, the only term contributing to the scale anomaly at order $m^0$ is the mass term. Recovering the factor $\frac{i}{16\pi^2}$ from the master integrals, we obtain the scale anomaly,
\begin{equation}
    \mathcal{A}=\sigma\frac{i}{24\pi^2}\tr\,F^2\, .
\end{equation}
The remaining trace is in gauge space only.

\bibliographystyle{JHEP}
\bibliography{biblio}
\end{document}